\documentclass[english,usenatbib]{mn2e}
\usepackage[T1]{fontenc}
\usepackage[latin9]{inputenc}
\usepackage{verbatim}
\usepackage{textcomp}
\usepackage{url}
\usepackage{relsize}
\usepackage{graphicx}
\usepackage[authoryear]{natbib}

\makeatletter

\DeclareRobustCommand{\lyxmathsym}[1]{\ifmmode\begingroup\def\b@ld{bold}
  \def\rmorbf##1{\ifx\math@version\b@ld\textbf{##1}\else\textrm{##1}\fi}
  \mathchoice{\hbox{\rmorbf{#1}}}{\hbox{\rmorbf{#1}}}
  {\hbox{\smaller[2]\rmorbf{#1}}}{\hbox{\smaller[3]\rmorbf{#1}}}
  \endgroup\else#1\fi}

\providecommand{\tabularnewline}{\\}

\def\runningtitle{2D Kinematics of 1.0$\ltapprox$z$\ltapprox$1.5 galaxies}
\def\runningauthor{M. Lemoine-Busserolle and F. Lamareille}

\def\kmsmpc{km s$^{-1}$ Mpc$^{-1}$}

\def\ergs{ergs s$^{-1}$}
\def\ergscm{ergs s$^{-1}$ cm$^{-2}$}
\def\ergshz{ergs s$^{-1}$ Hz$^{-1}$}
\def\msunyr{M$_{\odot}$ yr$^{-1}$}
\def\halpha{\ifmmode {\rm H{\alpha}} \else $\rm H{\alpha}$\fi}
\def\hbeta{\ifmmode {\rm H{\beta}} \else $\rm H{\beta}$\fi}
\def\oii{[O\,{\sc ii}] $\lambda\lambda$3726,3728}

\newcommand{\msun}{\,{\rm M_\odot}}

\newcommand{\ltapprox}{\raisebox{-0.5ex}{$\,\stackrel{<}{\scriptstyle\sim}\,$}}
\newcommand{\gtapprox}{\raisebox{-0.5ex}{$\,\stackrel{>}{\scriptstyle\sim}\,$}}

\newcommand{\whalf}{\hbox{$S_{0.5}$}}

\newcommand{\sigoned}{\hbox{$\sigma_{1D}$}}

\newcommand{\mnras}{MNRAS}%
\newcommand{\aap}{A\&A}%
\newcommand{\apj}{ApJ}%
\newcommand{\apjl}{ApJL}%
\newcommand{\araa}{ARA\&A}%
\newcommand{\pasp}{PASP}%
\newcommand{\nat}{Nature}%

\makeatother

\usepackage{babel}

\makeatother

\usepackage{babel}

\begin{document}

\title[\runningtitle]{2D Kinematics and physical properties of 1.0$\ltapprox$z$\ltapprox$1.5 star-forming galaxies%
\thanks{Based on data obtained with the European Southern Observatory Very
Large Telescope, Paranal, Chile, programs 075.A-0318 and 078.A-0177.%
}}

\author[\runningauthor]{M. Lemoine-Busserolle$^{1}$\thanks{mrlb@astro.ox.ac.uk} and F. Lamareille$^{2}$\\
$^{1}$\,Oxford Physics, University of Oxford, Keble Road, Oxford,
OX1\,3RH, UK\textbf{}\\
 $^{2}$\,Laboratoire d'Astrophysique de Toulouse-Tarbes, Universit\'e
de Toulouse, CNRS, 14 Avenue E. Belin, F-31400 Toulouse, France\\
}

\date{Accepted 2009 November 20.  Received 2009 November 12; in original form 2009 March 6}

\pagerange{\pageref{firstpage}--\pageref{lastpage}} \pubyear{2009}

\maketitle

\label{firstpage}

\begin{abstract}
We combined two-dimensional kinematic and morphology information on the \halpha\ emission, obtained using near-infrared integral field spectroscopy, with
broad-band photometry to investigate the dynamical structure and the physical properties of a sample of ten late-type galaxies at 1.0$\ltapprox$z$\ltapprox$1.5.
Their star formation rate ranges from $\sim$4 to $\sim$400 \msunyr with a mean value of $\sim$80 \msunyr. We found that three of these objects are undergoing a
strong burst of star formation. The sample displays a range of kinematical types which include one merger, one face-on galaxy, and eight objects showing evidence
of rotation. Among these eight objects, half are rotation-dominated galaxies, while the rest are dispersion-dominated. We found also that two galaxies out of the
rotation-dominated galaxies are pure rotationally supported disks. They achieve a maximum velocity of $\sim$ 180-290 km s$^{-1}$ within $\sim$ 0.5-1 kpc, similar
to local spirals with thin disks. Regarding the perturbed rotation and the dispersion-dominated galaxies, they display a plateau velocity range of 105-257 km
s$^{-1}$, which is certainly underestimated due to beam smearing. However, their plateau radii (4.5-10.8 kpc) derived from our rotating disk model are
significantly higher than those derived for pure rotating disks and local spiral galaxies. The galaxies of our sample have relatively young stellar populations
($\ltapprox$ 1.5 Gyr) and possess a range of stellar mass of 0.6-5 $\times 10^{10}\, M_{\odot}$. In addition, most of them have not yet converted the majority of
their gas into stars (six galaxies have their gas fraction $>$50 per cent). Therefore, those of them which already have a stable disk will probably have their
final stellar mass similar to the present-day spirals, to which these rotating systems can be seen as precursors. We conclude our study by investigating the
stellar mass Tully-Fisher relation at 1.2$\ltapprox$z$\ltapprox$1.5.
\end{abstract}
\section{Introduction}
In the last four years, observations by using integral field spectroscopy (IFS) of individual star-forming galaxies have revealed that galaxies at
$z\gtapprox1.5$ show a large variety of kinematic and dynamical properties. Recent studies at $z\sim 1.5-1.6$
\citep{Wright:2007ApJ...658...78W,Wright:2009ApJ...699..421W,Bournaud:2008A&A...486..741B} have confirmed that in addition of the evidence of organized
rotation non-negligible random motions are also detected at this redshift. In addition, it appears that large and massive disks with strong star
formation already exist at $z\gtapprox1.5$ \citep{Forster:2006ApJ...645.1062F,Genzel:2006Natur.442..786G,Genzel:2008ApJ...687...59G}. However more
numerous studies at $z\sim 2 - 3$
\citep{Genzel:2006Natur.442..786G,Forster:2006ApJ...645.1062F,2007ApJ...669..929L,2007ApJ...671..303B,Genzel:2008ApJ...687...59G,2008A&A...479...67N,2008A&A...488...99V,2009arXiv0901.2930L}
have found an increase in non-circular motions relatives to lower redshift samples. Indeed most of these high-redshift galaxies have high velocity
dispersions. Even when a large-scale velocity gradient throughout the galaxy is detected, the value of the ratio of $V/\sigma$ is relatively low ($<
10$) suggesting that it is unlikely for these galaxies to have a dynamically cold rotating disk of ionised gas similar to the local spiral galaxies
($V/\sigma \sim 15-20$ \citep{Dib:2006ApJ...638..797D}). Most of these `heated disks' appear also to be extremely rich in gas with evidence of high
turbulent star formation\citep{Lemoine-Busserolle:2009}. Another significant difference in the properties of some of these high-redshift objects in
comparison to the local Universe is the increase in the irregular and asymmetric shape of these galaxies. Although some of these irregular objects may
be associated with mergers, the morphology and kinematics of the majority of them are incompatible with being ongoing mergers
\citep{Shapiro:2008ApJ...682..231S} and suggest that they are rotationally supported
\citep{Genzel:2006Natur.442..786G,Forster:2006ApJ...645.1062F,DebraElmegreen07,Genzel:2008ApJ...687...59G}. These so-called `chain-galaxies' have the
appearance of giant highly clumpy disks. The objects can be kpc wide and as massive as $10^9 \msun$ \citep{Elmegreen05,Elmegreen09}.
\citet{Elmegreenetal2005} was the first to suggest that they could be the progenitor of $z \sim 1$ spirals. Although the origin of the `heated disks'
remains uncertain and highly debated, recent theoretical studies \citep{Dekel09a,Dekel09b} propose a bimodality in galaxy type by $z \sim 3$ with clumpy
star-forming disks and spheroid-dominated galaxies with low star formation rates (SFRs). At $z \ltapprox 1$, the disks should be stabilized by the dominant
stellar disks and bulges, suggested that the redshift region of $1 < z < 2$ is a crucial step in the formation of massive disk galaxies. Nowadays
theoretical studies try to understand the formation and evolution of extended turbulent rotating disks, and observational studies, as the ones presented
in this paper, probing the kinematics properties (like observed morphology, dynamical mass, etc), the stellar population properties (age of the
population, stellar masses, etc), star forming rates, and the properties of the ionized gas (mass de gas and gas fraction) of the $1 < z < 2$ disks are
mandatory for setting constrains on galaxy formation and evolution. %
\begin{table*}
\caption{Table of observations}

\resizebox{18cm}{!}{ \begin{tabular}{cccccccccc}
\hline 
Galaxy & VVDS-ID  & RA (J2000)  & DEC (J2000)  & z (1)  & Field (2)  & Grating  & Exp. Time  & Seeing (3)  & Run ID \tabularnewline
\hline 
VVDS-1235 & 020461235 & 02:26:47.110  & -04:23:55.71  & 1.0351  & VVDS-02h  & J  & 1.11h  & 0\farcs.80  & 078.A-0177(A) \tabularnewline
VVDS-2331 & 020182331 & 02:26:44.260  & -04:35:51.89  & 1.2286  & VVDS-02h  & H  & 3h  & 0\farcs.93  & 078.A-0177(A) \tabularnewline
VVDS-6913 & 220596913 & 22:14:29.184  & +00:22:18.89  & 1.2667  & VVDS-22h  & H  & 1.75h  & 0\farcs.47  & 075.A-0318(A) \tabularnewline
VVDS-5726 & 220015726 & 22:15:42.455  & +00:29:03.59  & 1.3091  & VVDS-22h  & H  & 2h  & 0\farcs.55  & 075.A-0318(A) \tabularnewline
VVDS-4252 & 220014252 & 22:17:45.690  & +00:28:39.47  & 1.3097  & VVDS-22h  & H  & 2h  & 0\farcs.61  & 075.A-0318(A) \tabularnewline
VVDS-4103 & 220544103 & 22:15:25.708  & +00:06:39.53  & 1.3970  & VVDS-22h  & J \& H  & 1h (J); 1h (H)  & 0\farcs.69  & 075.A-0318(A) \tabularnewline
VVDS-4167 & 220584167 & 22:15:23.038  & +00:18:47.01  & 1.4637  & VVDS-22h  & J \& H  & 1h (J); 1.75h (H)  & 0.\farcs77  & 075.A-0318(A) \tabularnewline
VVDS-7106 & 020147106 & 02:26:45.386  & -04:40:47.39  & 1.5174  & VVDS-02h  & H  & 2h  & 0\farcs.89  & 075.A-0318(A) \tabularnewline
VVDS-6027 & 020116027 & 02:25:51.133  & -04:45:04.48  & 1.5259  & VVDS-02h  & H  & 1.67h  & 0\farcs.65  & 075.A-0318(A) \tabularnewline
VVDS-1328 & 020261328 & 02:27:11.049  & -04:25:31.60  & 1.5291  & VVDS-02h  & H  & 1h  & 0\farcs.61  & 075.A-0318(A) \tabularnewline
\hline
\end{tabular}}

\raggedright{}The columns are as follows: (1) redshift estimated from
optical spectrum obtained with VIMOS, (2) VVDS-22h wide field ($17.5\leq I_{AB}\leq22.5$)
and VVDS-02h deep field ($17.5\leq I_{AB}\leq24.0$), (3) median seeing  estimated from the PSF stars (one taken per hour of observation).
\label{runs} 
\end{table*}
 
At lower redshifts, the velocity fields of spiral/rotating disk galaxies have been used to place important
constraints on total masses and hence on dark matter halo masses \citep{Conselice:2005ApJ...628..160C}. These
rotating disks produce a Tully-Fisher relationship (the scaling law between total dynamical mass and the luminous
stellar mass) which has apparently not evolved in slope and scatter since $z\approx 0.6$
\citep{Puech:2006A&A...455..119P,Puech:2008A&A...484..173P}. Therefore the large scatter found in previously
reported Tully-Fisher relationships at moderate redshifts is produced by galaxies with perturbed rotation or
complex kinematics. The importance of the role of the non-ordered motion through the gas velocity dispersion for
galaxies showing a perturbed rotation was investigated by \citet{Weiner:2006ApJ...653.1027W} and
\citet{Kassin:2007ApJ...660L..35K}. They defined a new tracer of galaxy-dark halo potential, which combines
dynamical support from the rotation with that of the non-ordered motion. The new stellar mass Tully-Fisher
relation shows no detectable evolution neither in slope or of its intercept up to z$\sim$1.2\textbf{
}\citep{Kassin:2007ApJ...660L..35K}. Two-dimensional velocity fields allow rotation curves to be deduced in a
more robust manner than slit spectra\textbf{ }\citep{Weiner:2006ApJ...653.1027W}. IFS allows pure
rotationally-supported disks to be distinguished from other dynamically-disturbed galaxies, which include minor
or major mergers, merger remnants and/or inflow/outflows.

In this paper, we explore the kinematic properties of the
ionized gas and star formation rates of a sample of ten galaxies at
$1<z<1.5$ selected in the VIMOS VLT Deep Survey (VVDS), using integral field unit (IFU) $H$-band and $J$-band
spectroscopy with VLT/SINFONI. The results presented here are part
of a study to investigate kinematics and physical properties of unbiased samples representative of the global
intermediate and high-$z$ population. This is the second
paper of three articles based on data obtained during two observing
runs with SINFONI. From the two other companion papers, one \citep{Queyrel2009A&A...506..681Q} diskusses
the chemical properties of the intermediate redshift sample presented
here and the other presents the results
found on the $z\sim3$ galaxy sample \citep{Lemoine-Busserolle:2009}.

In Section \ref{Obs} we describe our sample, observational strategy
and the data reduction techniques. In Section \ref{sprectro_prop}
we address the properties of the nebular emission and also the nature of
the stellar population of the galaxies, using broad-band photometry
to constrain properties such as stellar mass, age and SFRs. In Section \ref{kine_prop} we investigate the kinematic structure
and dynamical properties inferred from IFS.
In Section \ref{nature} we explore the nature of these 
star-forming galaxies and in Section \ref{relations} we diskuss various relations between the properties of the gas and those of the stellar population  and compare our findings with previous studies of the kinematics
 of intermediate and high-redshift galaxies. Finally in Section \ref{summary} we summarize
our results and diskuss their implications.

We assume a cosmology with $\Omega_{0}=0.3$, $\Lambda=0.7$ and $H_{0}$
= 70 \kmsmpc throughout, and all magnitudes are on the $AB$ system
\citep{Oke:1983ApJ...266..713O}.
\begin{table*}
\caption{Nebular emission properties}

\resizebox{18cm}{!}{ \begin{tabular}{cccccccccc}
\hline 
Galaxy  & $z_{\halpha}$  & F(H$\alpha$) (1)  & L(H$\alpha$) (2)  & $SFR_{H\alpha}^{0}$ (3)  & E(B-V)$_{gas}$ (4)  & $SFR_{corr}$ (5) & $\sigma_{1D}$ (6) & $M_{gas}$ (7)  & $\mu$ (8)\tabularnewline
\hline 
VVDS-1235  & $1.0352$  & $20.1\pm2$  & $11.4\pm1$  & $5.3\pm0.5$  & $0.17\pm0.02$  & $8.8\pm1$  & $74\pm9$ & $8.4\pm0.7$  & $0.58\pm0.08$\tabularnewline
VVDS-2331  & $1.2286$  & $45\pm8$  & $39.1\pm7$  & $18\pm3$  & $0.28\pm0.03$  & $42\pm8$  & $94\pm21$ & $31\pm4$  & $0.74\pm0.06$\tabularnewline
VVDS-6913  & $1.2660$  & $29.7\pm3$  & $27.8\pm3$  & $12.8\pm1$  & $0.09\pm0.02$  & $17\pm2$  & $113\pm18$ & $22\pm2$  & $0.34\pm0.09$\tabularnewline
VVDS-5726  & $1.2927$  & $23.4\pm0.3$  & $23.8\pm0.3$  & $10.9\pm0.1$  & $0.24\pm0.02$  & $23\pm2$  & $128\pm2$ & $19\pm1$  & $0.28\pm0.07$\tabularnewline
VVDS-4252  & $1.3099$  & $201\pm54$  & $205\pm55$  & $94\pm25$  & $0.36\pm0.04$  & $280\pm83$  & $129\pm40$ & $143\pm30$  & $0.87\pm0.04$\tabularnewline
VVDS-4103  & $1.3966$  & $208\pm49$  & $248\pm59$  & $114\pm27$  & $0.42\pm0.04$  & $412\pm108$  & $107\pm57$ & $208\pm39$  & $0.93\pm0.02$\tabularnewline
VVDS-4167  & $1.4656$  & $33.3\pm1$  & $44.6\pm1$  & $20.5\pm0.6$  & $0.10\pm0.01$  & $28\pm1$  & $141\pm6$ & $34.1\pm0.8$  & $0.58\pm0.08$\tabularnewline
VVDS-7106  & $1.5194$  & $21.2\pm2$  & $31.1\pm3$  & $14.3\pm1$  & $0.05\pm0.01$ & $16\pm2$ & $82\pm7$ & $16\pm1$  & $0.67\pm0.06$\tabularnewline
VVDS-6027  & $1.5301$  & $5.63\pm0.6$  & $8.4\pm0.9$  & $3.85\pm0.4$  & $0.00\pm0.02$  & $3.9\pm0.4$ & $42\pm7$ & $5.5\pm0.4$  & $0.33\pm0.08$\tabularnewline
VVDS-1328  & $1.5289$  & $5.48\pm0.6$  & $8.2\pm0.9$  & $3.76\pm0.4$  & $0.01\pm0.02$ & $3.9\pm0.5$ & $78\pm11$ & $5.7\pm0.5$  & $0.34\pm0.12$\tabularnewline
\hline
\end{tabular}}

The columns are as follows: (1) Flux($\times10^{-17}$ \ergscm),
(2) Luminosity($\times10^{+41}$ \ergs), (3) raw SFR derived from
\halpha-line (\msunyr), (4) reddening suffered by the ionized gas,
(5) dereddened SFR, (6) global velocity dispersion derived from \halpha-line
(km/s), (7) total mass of ionized gas ($10^{9}\, M_{\odot}$), (8)
gas mass fraction. \label{tabline} 
\end{table*}
\section{Data and Observations}
\label{Obs}
\subsection{Sample Selection}
\label{sample} 
The 10 galaxies, presented in this paper, were selected in the VVDS-02h {}``deep'' field ($17.5\leq I_{AB}\leq24.0$) of the VVDS
\citep{LeFevre:2005A&A...439..845L} and in the VVDS-22h {}``wide'' field ($17.5\leq I_{AB}\leq22.5$; \citealt{Garilli:2008A&A...486..683G}). The objects
of this sample have spectroscopic redshifts, derived from VLT/VIMOS spectroscopy, between $z\sim1.03$ and $z\sim1.53$ (see Table\,\ref{runs}). Their
redshifts are such that the expected wavelengths of the rest-optical H$\alpha$ and {[}N\,II{]} emission lines would be clear of bright OH sky lines in
the near-infrared (NIR). These targets were mainly selected on the basis of their measured intensity of \oii\ emission lines in the VIMOS spectra. We
chose objects showing the strongest {[}OII{]}3727 emission lines (rest-frame equivalent width (EW) $>50$ \AA\ and flux $>5\times10^{-17}$
ergs$^{-1}$cm$^{-2}$) and having a photometric spectral energy distribution (SED) corresponding to late-type star-forming galaxies.
\subsection{SINFONI observations}
The NIR spectroscopic observations were acquired with the 3D spectrograph SINFONI at ESO-VLT during two 4-nights runs, on 2005 September 5-8 (ESO run
75.A-0318) and 2006 November 12-15 (ESO run 78.A-0177). SINFONI was used in its seeing-limited mode, with the 0.25\arcsec\ pixel scale leading to a
field-of-view of 8$\times$8 $arcsec^{2}$, and the $H$ grism (1.447 - 1.847 $\mu m$) providing a spectral resolution $R \sim 4000$. One galaxy,
VVDS-1235, has been observed with the $J$ grism (1.098 - 1.399 $\mu m$), where the detection of the \halpha\ emission line is expected. Three galaxies
(VVDS-1235, VVDS-4103 and VVDS-4167) have also been observed with the $J$ grism (1.098 - 1.399 $\mu m$). Conditions were photometric and the median
seeing for each objects, estimated on 2D imaging of point spread function (PSF) stars taken each hour of object observation, is indicated in
Table\,\ref{runs}. Observing strategy and data reduction have been described in \citet{Lemoine-Busserolle:2009}.
\section{Spectrophotometry properties} 
\label{sprectro_prop}
\subsection{Nebular integrated emission properties}
\subsubsection{One-dimensional spectral properties of \halpha\ emission }
\begin{figure*}
\begin{centering}
\includegraphics[width=1\columnwidth]{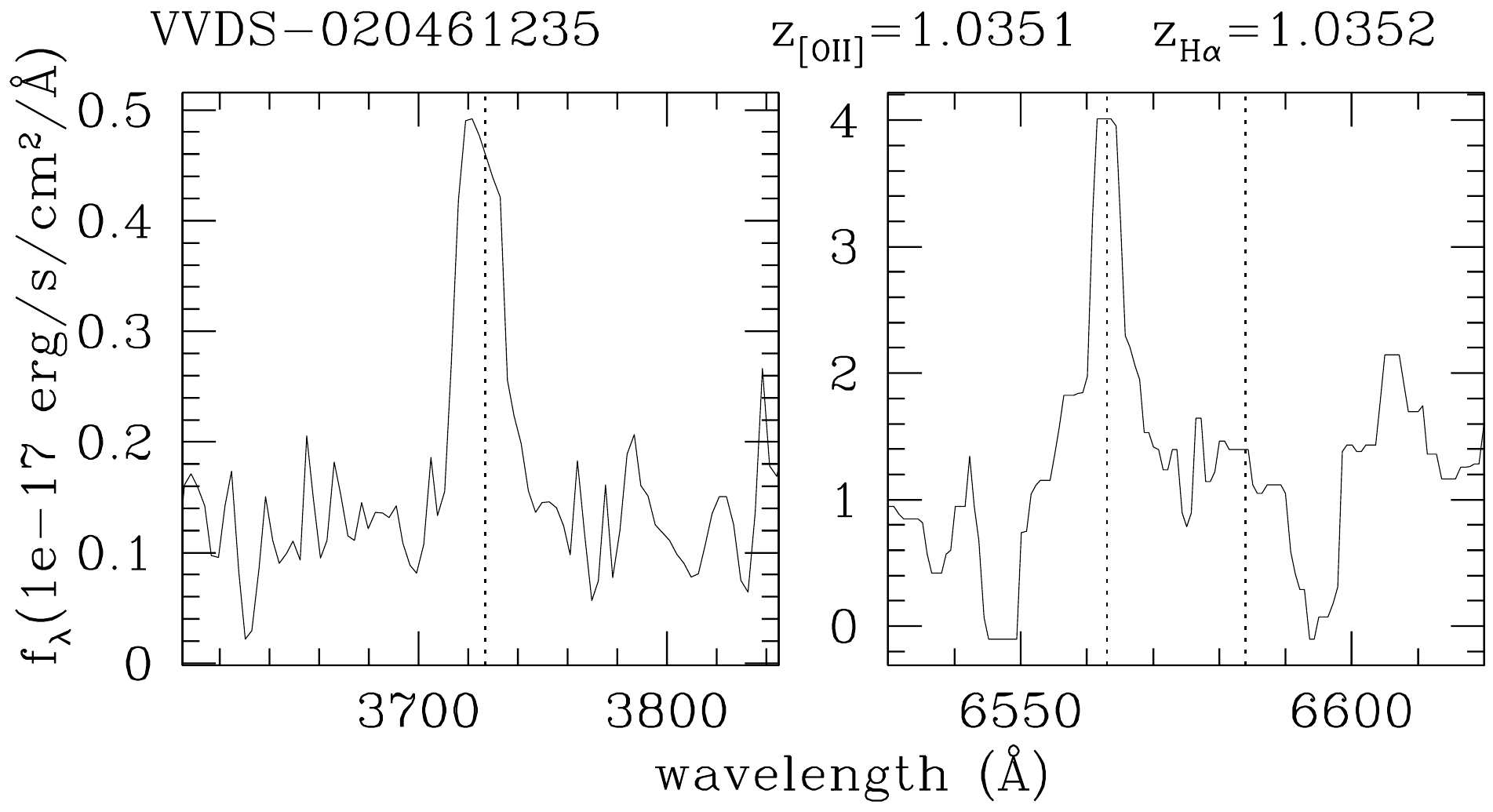} \includegraphics[width=1\columnwidth]{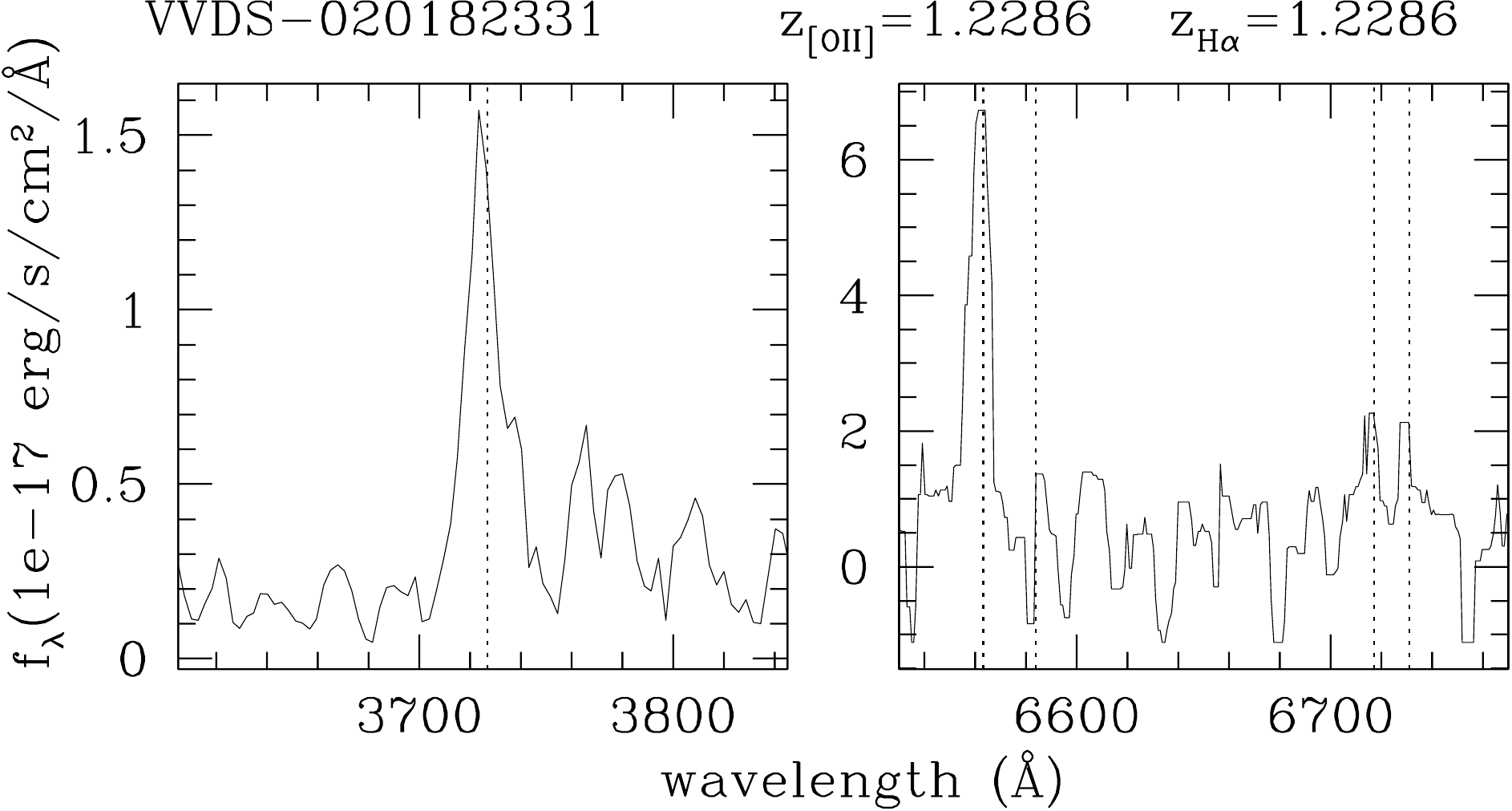}
\includegraphics[width=1\columnwidth]{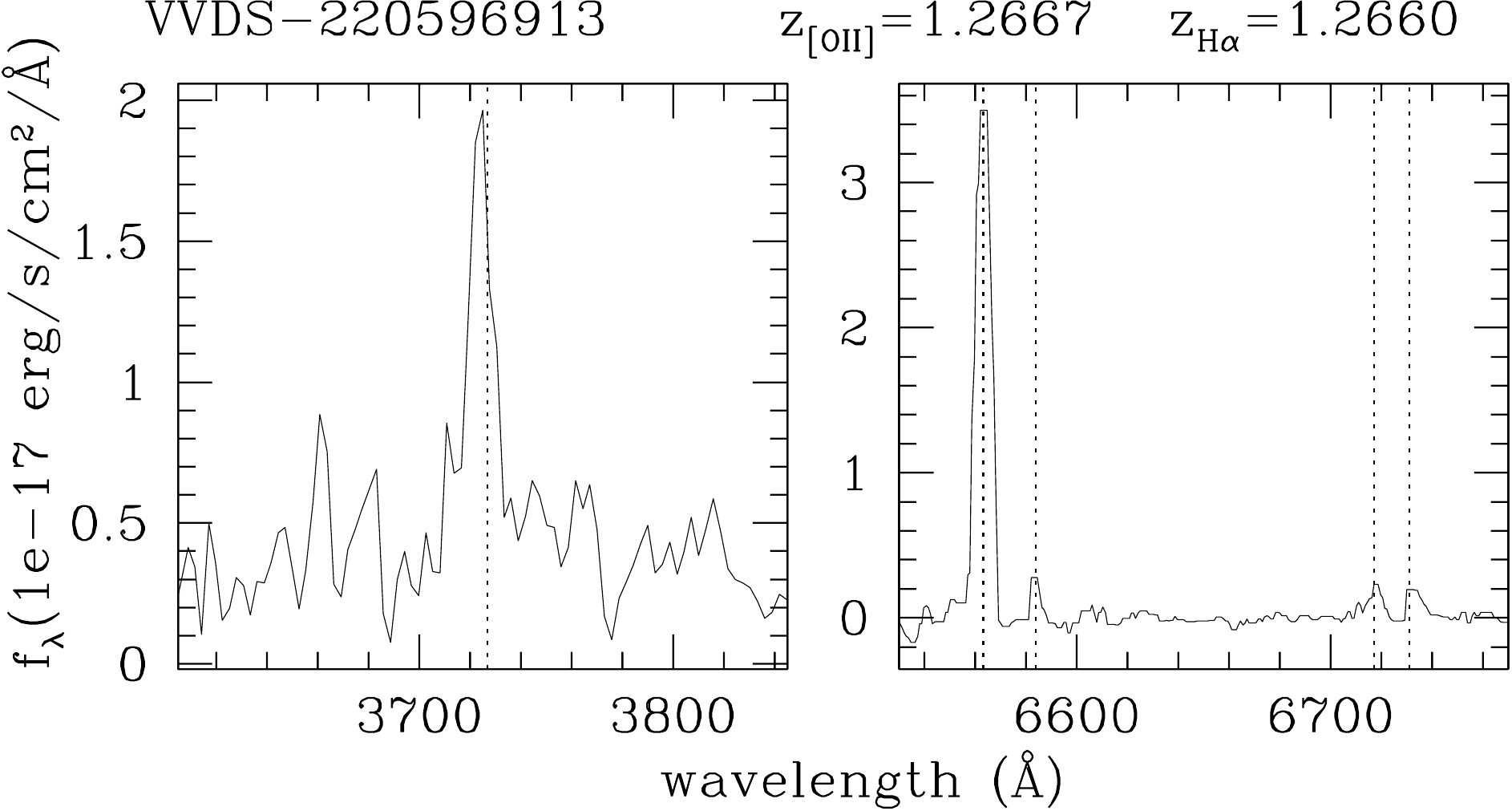} \includegraphics[width=1\columnwidth]{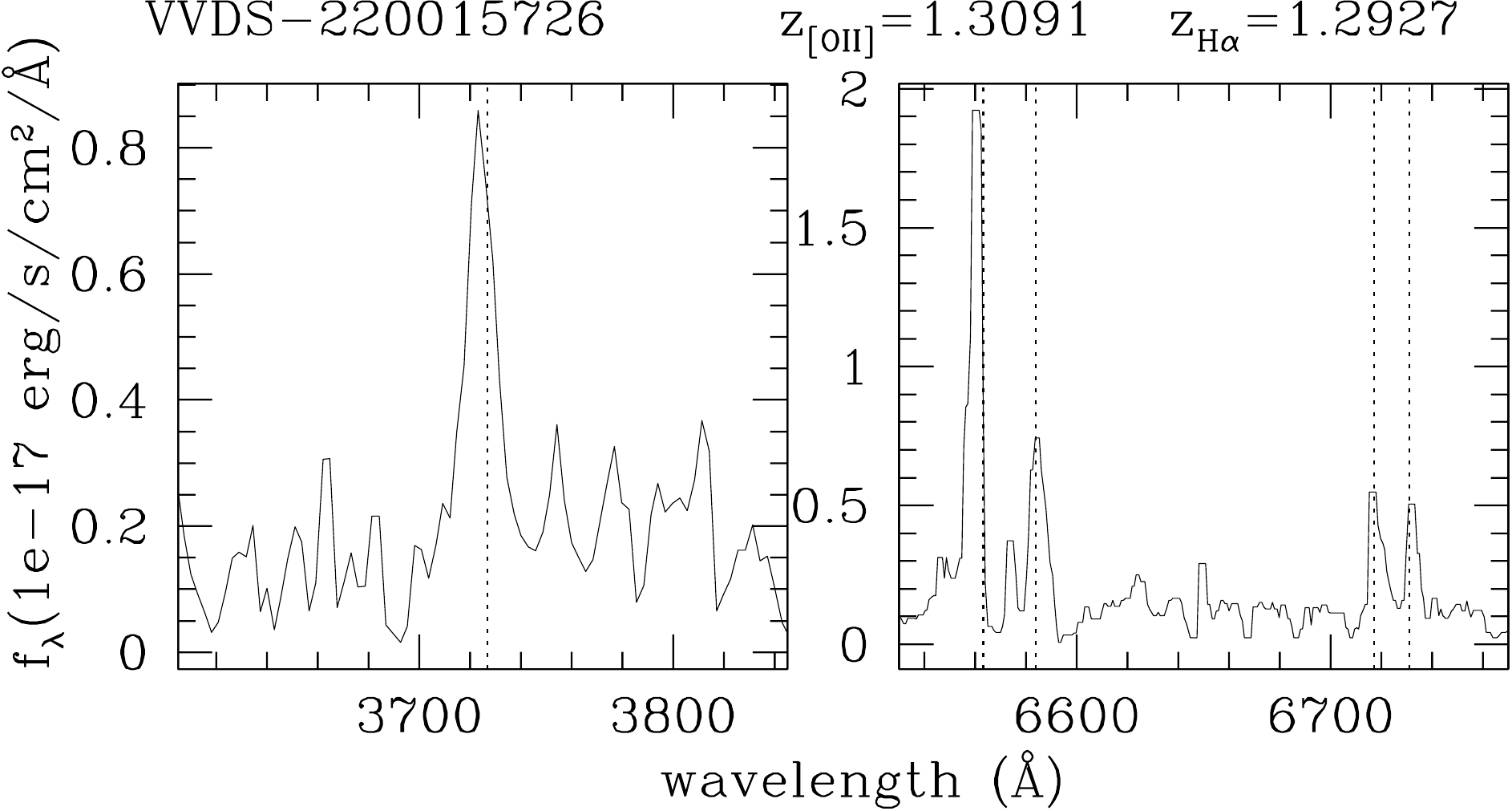}
\includegraphics[width=1\columnwidth]{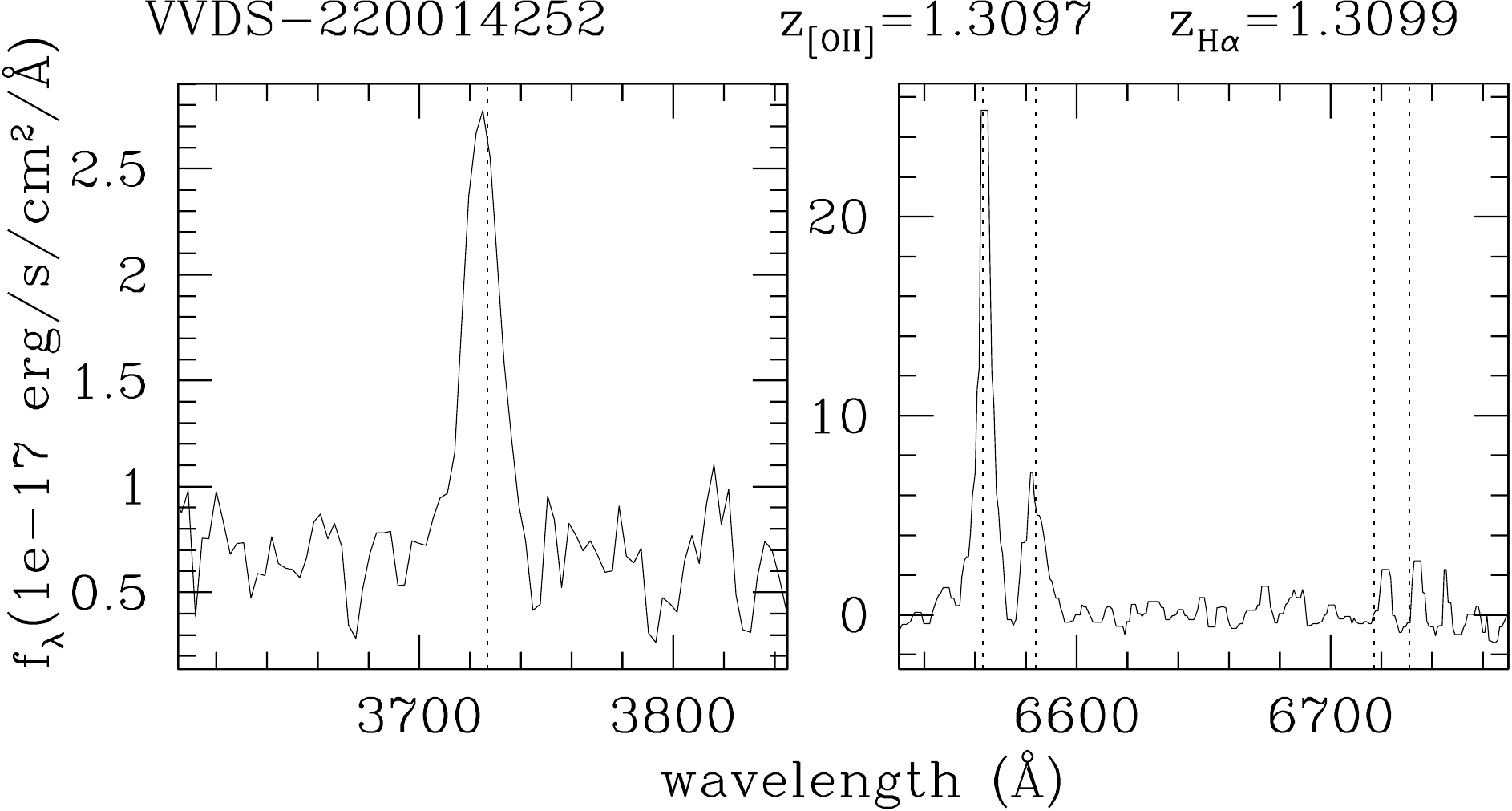} \includegraphics[width=1\columnwidth]{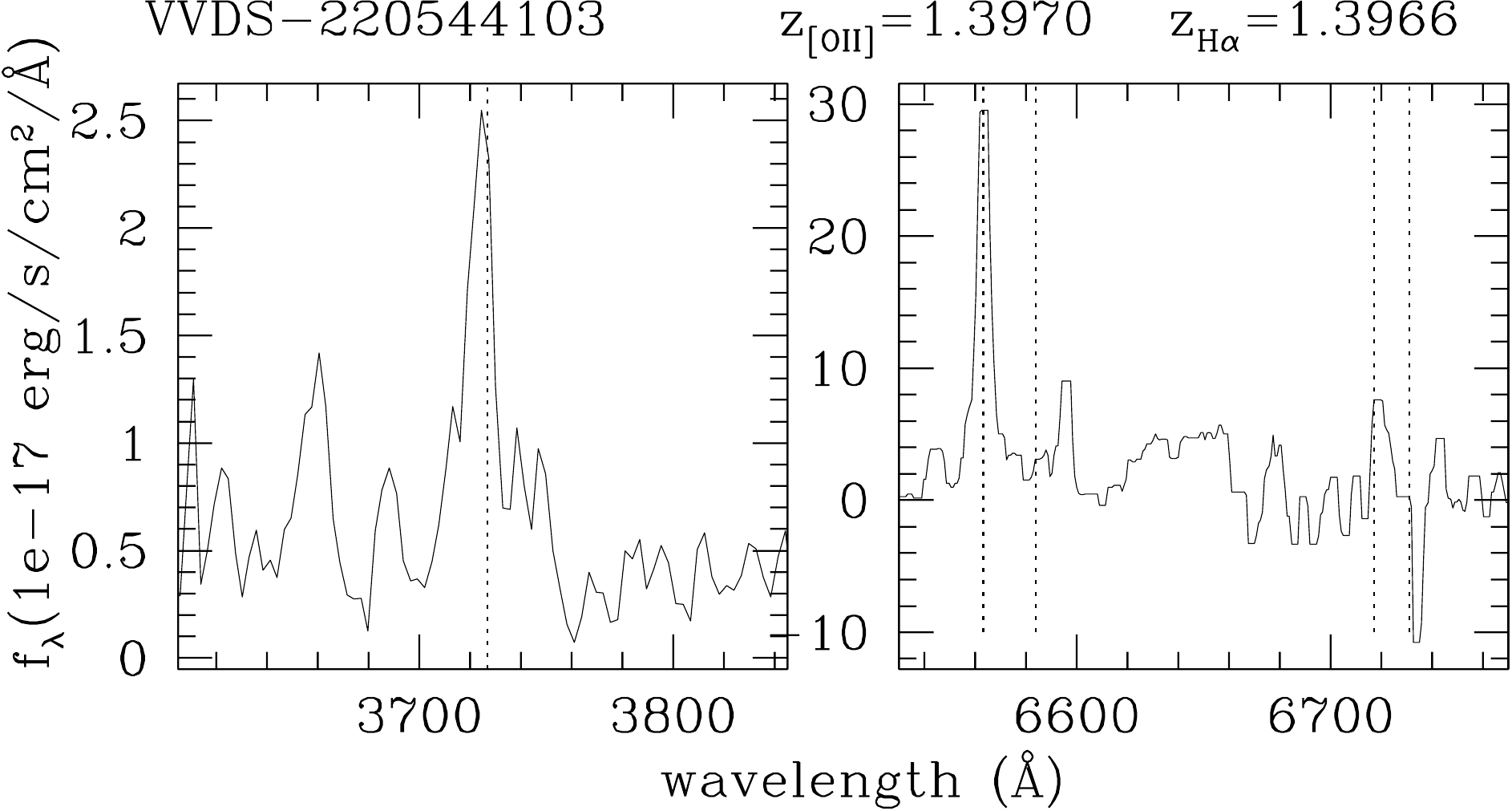}
\includegraphics[width=1\columnwidth]{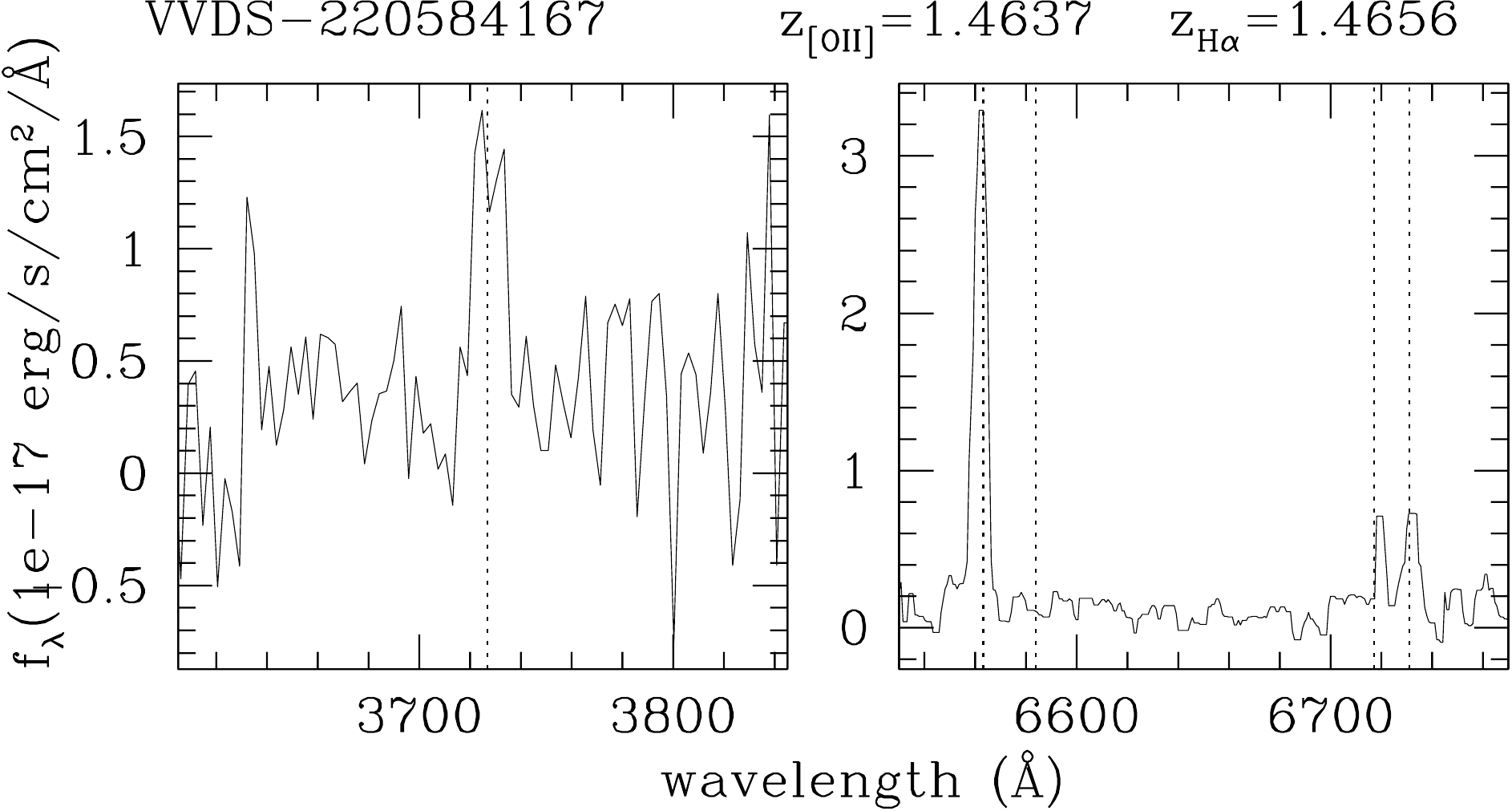} \includegraphics[width=1\columnwidth]{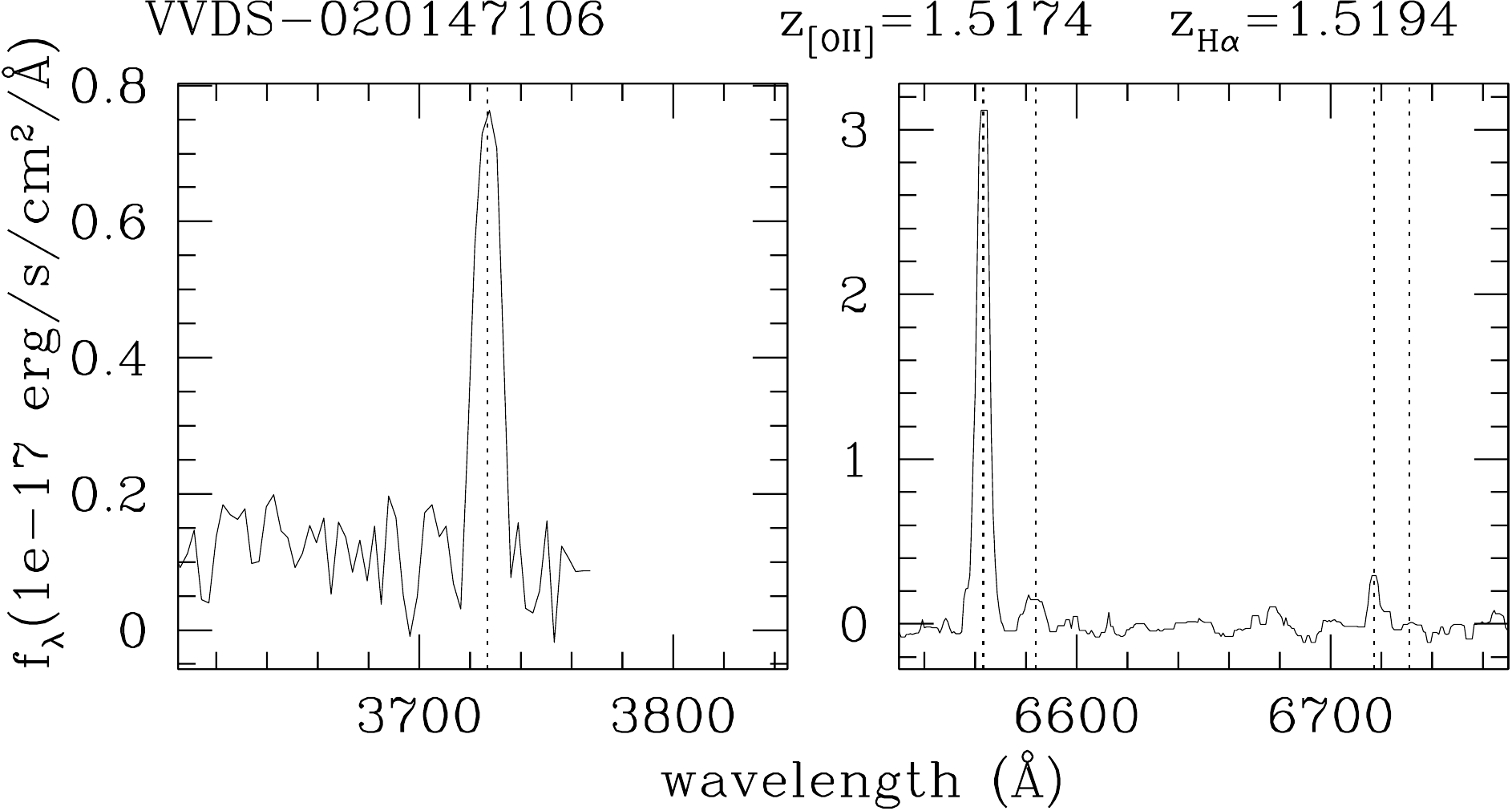}
\includegraphics[width=1\columnwidth]{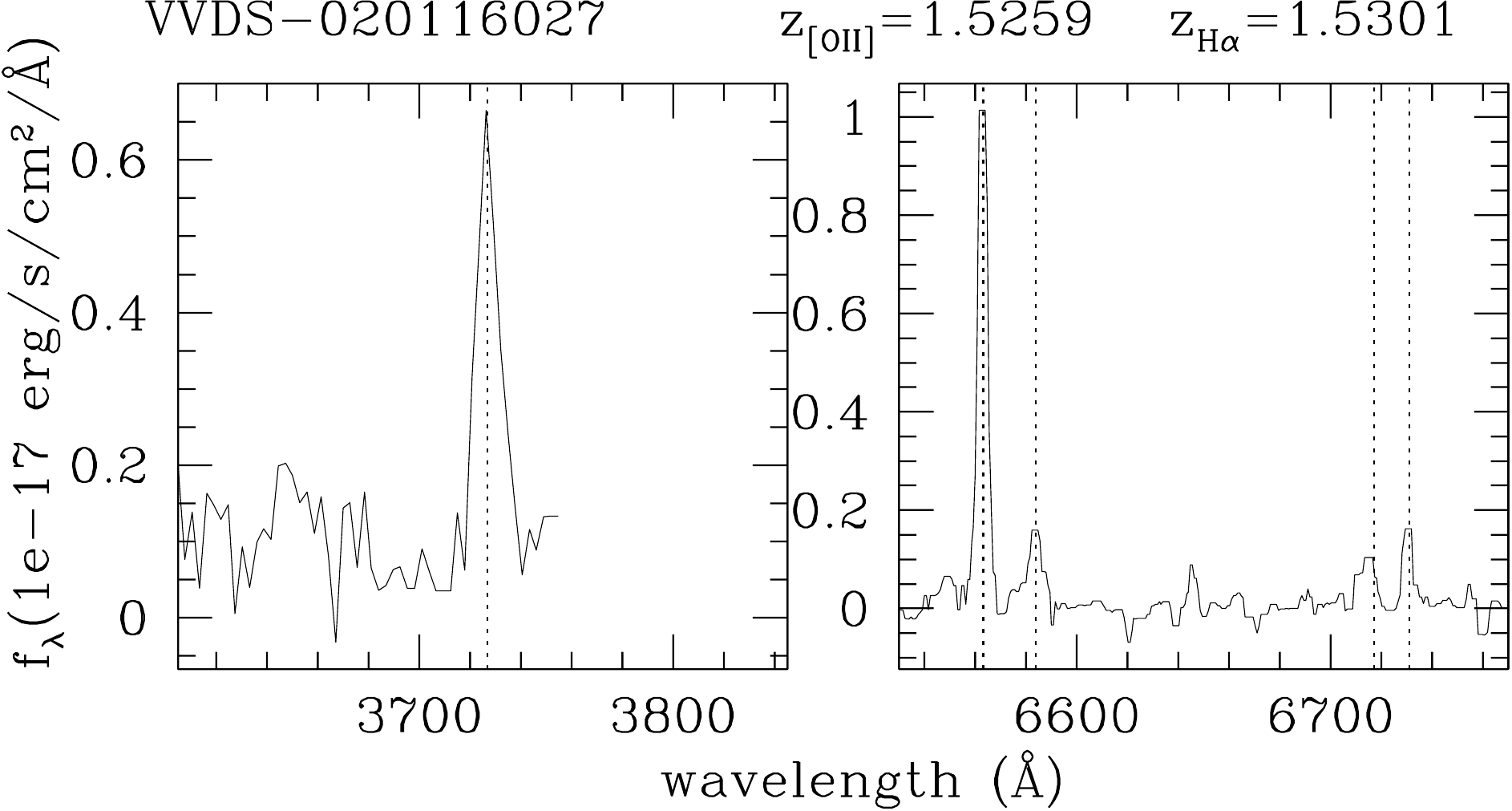} \includegraphics[width=1\columnwidth]{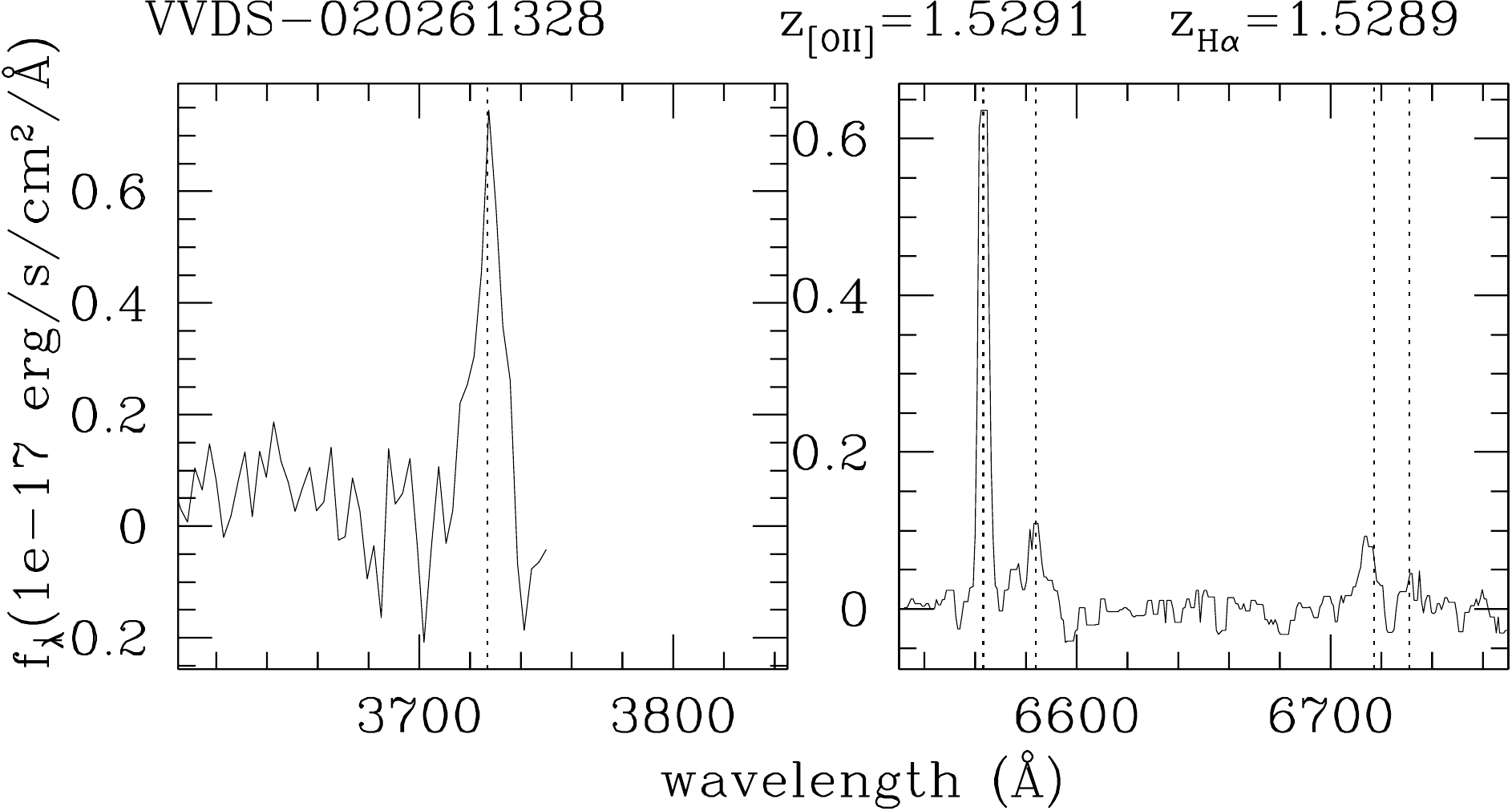} 
\par\end{centering}

\caption{One-dimensional rest-frame spectra for the 10 galaxies of our sample obtained with VIMOS (left) and with
SINFONI (right), in optical and in $H$-band (or $J$-band for VVDS-1235) respectively. The position of the following lines are
shown by a vertical line, from left to right: {[}OII{]}$\lambda$3727,
H$\alpha$, {[}NII{]}$\lambda$6584, {[}SII{]}$\lambda$6717, and
{[}SII{]}$\lambda$6731. The redshifts inferred from {[}OII{]}$\lambda$3727
and from H$\alpha$ lines are indicated in the title of each plots.}

\label{fig1D}
\end{figure*}

Integrated rest-frame optical one-dimensional spectra were extracted from the SINFONI data cube for each object. From these 1D spectra, we measured line
fluxes and linewidths of the integrated \halpha\ line emission (see Fig.\ref{fig1D}). Using the IRAF task {}`splot', we derived the systemic redshift
$z_{\halpha}$, the total \halpha\ flux and the global velocity dispersion $\sigma_{1D}$ (corrected for the instrumental resolution of $\sim$6.5 \AA\ in
the $J$ band and $\sim$6.8 \AA\ in the $H$ band). Table~\ref{tabline} list the properties of the nebular emission of the 10 galaxies of our sample. It
is interesting to note that all our targets, except VVDS-6027, have a value of $\sigma_{1D} > 70$ km s$^{-1}$.
%
\subsubsection{Star Formation rates from \halpha\ emission}
The H$\alpha$ nebular recombination line is a direct probe of the young, massive stellar population and therefore provides a nearly instantaneous (i.e.
averaged on the last ten million years) measure of the SFR. The H$\alpha$ line is moreover not strongly affected by dust extinction. We have calculated
$SFR_{H\alpha}$ following the \citet{kennicutt:1998ARA&A..36..189K} calibration, re-normalized to \citet{chabrier:2003PASP..115..763C} IMF:
\begin{equation}
SFR_{\halpha}(\msun\textup{ yr}^{-1})=4.6\times10^{-42}\; L(H\alpha)\:\:(\textup{erg s}^{-1}).
\end{equation}

Table~\ref{tabline} list the star formation rates deduced from the \halpha\ emission for the 10 galaxies of our sample. We found that the
non-dust-corrected $SFR_{\halpha}$ is between $\sim$4 $-$ 114 $\msun yr^{-1}$, with seven of our galaxies with a $SFR_{\halpha} >$ 10 $\msun yr^{-1}$.
These high SFRs are expected, taking into account our selection criteria for the [OII] flux (see section \ref{sample}).
\begin{table*}
\caption{Photometric data and Spectral energy distributions}

\resizebox{18cm}{!}{ \begin{tabular}{cccccccccccc}
\hline 
Galaxy & B$_{AB}$ & V$_{AB}$ & R$_{AB}$ & I$_{AB}$ & u$_{AB}$ & g$_{AB}$ & r$_{AB}$ & i$_{AB}$ & z$_{AB}$ & J$_{AB}$ & K$_{AB}$\tabularnewline
\hline 
VVDS-1235  & $24.1\pm0.06$  & $23.7\pm0.05$  & $23.4\pm0.05$  & $22.6\pm0.05$  & $24.1\pm0.03$  & $23.8\pm0.02$  & $23.5\pm0.02$  & $22.8\pm0.01$  & $22.7\pm0.04$  & $22.3\pm0.05$  & $21.8\pm0.06$ \tabularnewline
VVDS-2331  & $24.0\pm0.09$  & $23.6\pm0.07$  & $23.4\pm0.07$  & $22.7\pm0.06$  & $24.0\pm0.03$  & $23.6\pm0.02$  & $23.2\pm0.02$  & $22.9\pm0.01$  & $22.3\pm0.03$  & $22.3\pm0.10$$^{\star}$ & $21.6\pm0.11$ \tabularnewline
VVDS-6913  & -  & -  & -  & $21.7\pm0.14$  & $23.1\pm0.02$  & $22.6\pm0.01$  & $22.3\pm0.01$  & $22.1\pm0.01$  & $21.3\pm0.00$  & -  & - \tabularnewline
VVDS-5726  & -  & -  & -  & $22.4\pm0.17$  & $24.3\pm0.04$  & $23.6\pm0.02$  & $23.2\pm0.01$  & $22.7\pm0.01$  & $22.1\pm0.01$  & -  & - \tabularnewline
VVDS-4252  & $22.5\pm0.03$  & $22.4\pm0.02$  & $22.1\pm0.01$  & $22.0\pm0.17$  & $22.9\pm0.01$  & $22.4\pm0.01$  & $22.1\pm0.01$  & $22.0\pm0.00$  & $21.5\pm0.00$  & $21.5\pm0.08$$^{\star}$  & $21.0\pm0.08$$^{\star}$\tabularnewline
VVDS-4103  & -  & -  & -  & $22.3\pm0.17$  & $23.2\pm0.02$  & $22.7\pm0.01$  & $22.4\pm0.01$  & $22.3\pm0.01$  & $22.1\pm0.01$  & -  & - \tabularnewline
VVDS-4167  & -  & -  & -  & $21.9\pm0.10$  & -  & -  & -  & -  & -  & -  & $21.3\pm0.20$$^{\star}$ \tabularnewline
VVDS-7106  & $22.9\pm0.04$  & $22.8\pm0.04$  & $22.7\pm0.04$  & $22.5\pm0.04$  & $23.0\pm0.01$  & $22.7\pm0.01$  & $22.6\pm0.01$  & $22.5\pm0.01$  & $22.4\pm0.02$  & $22.0\pm0.15$$^{\star}$  & $22.1\pm0.15$$^{\star}$ \tabularnewline
VVDS-6027  & $23.6\pm0.07$  & $23.2\pm0.05$  & $23.1\pm0.05$  & $22.9\pm0.08$  & $24.0\pm0.04$  & $23.4\pm0.02$  & $23.3\pm0.02$  & $23.1\pm0.02$  & $22.8\pm0.05$  & $22.6\pm0.50$$^{\star}$  & $23.2\pm0.50$$^{\star}$ \tabularnewline
VVDS-1328  & $24.3\pm0.07$  & $24.1\pm0.06$  & $24.2\pm0.08$  & $23.9\pm0.12$  & $24.2\pm0.04$  & $23.8\pm0.03$  & $23.6\pm0.03$  & $23.5\pm0.03$  & $23.6\pm0.10$  & $23.1\pm0.20$ & $22.2\pm8.03$\tabularnewline
\hline
\end{tabular}}

\begin{raggedright}
\par\end{raggedright}

\raggedright{}$^{\star}$: magnitudes from the UKIDSS survey.\label{tabmag} 
\end{table*}
\subsection{Stellar population properties}
\subsubsection{Spectral Energy Distribution and best-fit models}
We derived the spectral energy distributions (SED) of our 10 galaxies from broad-band photometry. The magnitudes in $BVRI$ filters obtained with the
CFH12k camera on CFHT \citep{McCracken:2003A&A...410...17M,LeFevre:2004A&A...417..839L} were extracted from the VVDS catalog
\url{http://cencosw.oamp.fr}.
\citep{Iovino:2005A&A...442..423I,Temporin:2008A&A...482...81T}, and magnitudes in $ugriz$ filters from the CFHTLS T003 release
\citep{Ilbert:2006A&A...457..841I}. Finally, we have also used magnitudes in $J$ and $K$ bands from the UKIDSS survey
\citep{Lawrence:2007MNRAS.379.1599L}. All the photometric data are presented in Table~\ref{tabmag}.
\begin{table*}
\caption{Stellar population properties}
 	\begin{tabular}{ccccccc}
\hline 
Galaxy  & $M_{\star}$ (1)  & $L_{1500}$ (2)  & $SFR_{UV}^{0}$ (3)  & $SFR_{sed}$ (4)  & E(B-V)$_{sed}$ (5)  & age (6)\tabularnewline
\hline 
VVDS-1235  & $6\pm2$  & $2.3\pm0.2$  & $1.8\pm0.2$  & $11\pm5$  & $0.28\pm0.12$  & $0.68\pm0.3$\tabularnewline
VVDS-2331  & $11\pm3$  & $3.7\pm0.1$  & $3\pm0.1$  & $27\pm14$  & $0.32\pm0.11$  & $0.54\pm0.3$\tabularnewline
VVDS-6913  & $43\pm16$  & $8.7\pm0.4$  & $7.1\pm0.4$  & $49\pm22$  & $0.28\pm0.10$  & $0.96\pm0.5$\tabularnewline
VVDS-5726  & $50\pm17$  & $3.1\pm0.4$  & $2.4\pm0.4$  & $42\pm28$  & $0.37\pm0.13$  & $1.26\pm0.7$\tabularnewline
VVDS-4252  & $21\pm7$  & $11.2\pm0.3$  & $9.4\pm0.2$  & $73\pm26$  & $0.28\pm0.08$  & $0.40\pm0.2$\tabularnewline
VVDS-4103  & $16\pm4$  & $9.4\pm0.5$  & $7.7\pm0.4$  & $91\pm49$  & $0.28\pm0.03$  & $1.75\pm1.6$\tabularnewline
VVDS-4167  & $25\pm8$  & $12.5\pm0.5$  & $10.6\pm0.4$  & $129\pm46$  & $0.29\pm0.07$  & $0.29\pm0.1$\tabularnewline
VVDS-7106  & $8\pm2$  & $12.9\pm0.1$  & $10.6\pm0.1$  & $44\pm11$  & $0.16\pm0.04$  & $0.27\pm0.1$\tabularnewline
VVDS-6027  & $11\pm4$  & $5.1\pm0.2$  & $4.1\pm0.2$  & $33\pm12$  & $0.24\pm0.08$  & $0.46\pm0.3$\tabularnewline
VVDS-1328  & $11\pm6$  & $4.4\pm0.2$  & $3.5\pm0.2$  & $14\pm6$  & $0.19\pm0.11$  & $0.88\pm0.6$\tabularnewline
\hline
	\end{tabular}
	
	\begin{raggedright}
	\par\end{raggedright}

\raggedright{}The columns are as follows: (1) Stellar mass ($10^{9}\, M_{\odot}$),
(2) Luminosity at $1500$\AA{} ($\times10^{+28}$ \ergshz), (3)
raw SFR derived from UV luminosity (\msunyr), (4) deredenned SFR
derived from the full SED (averaged in the last $10^{8}$ years),
(5) reddening suffered by the stars, (6) age of the oldest stellar populations
(Gyr). 
\label{tabsed} 
\end{table*}

\begin{figure*}
\begin{centering}
\includegraphics[width=1\columnwidth]{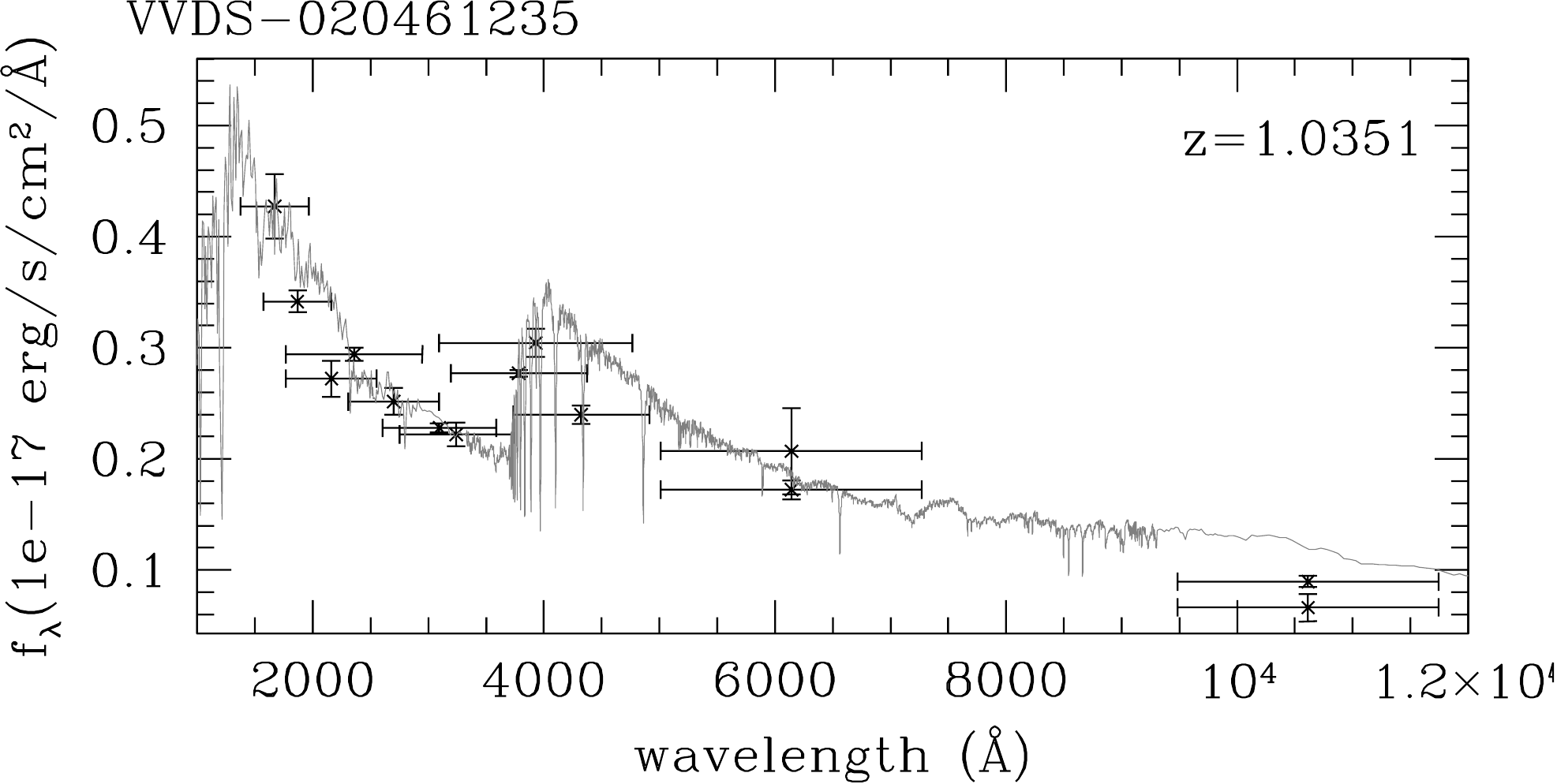} \includegraphics[width=1\columnwidth]{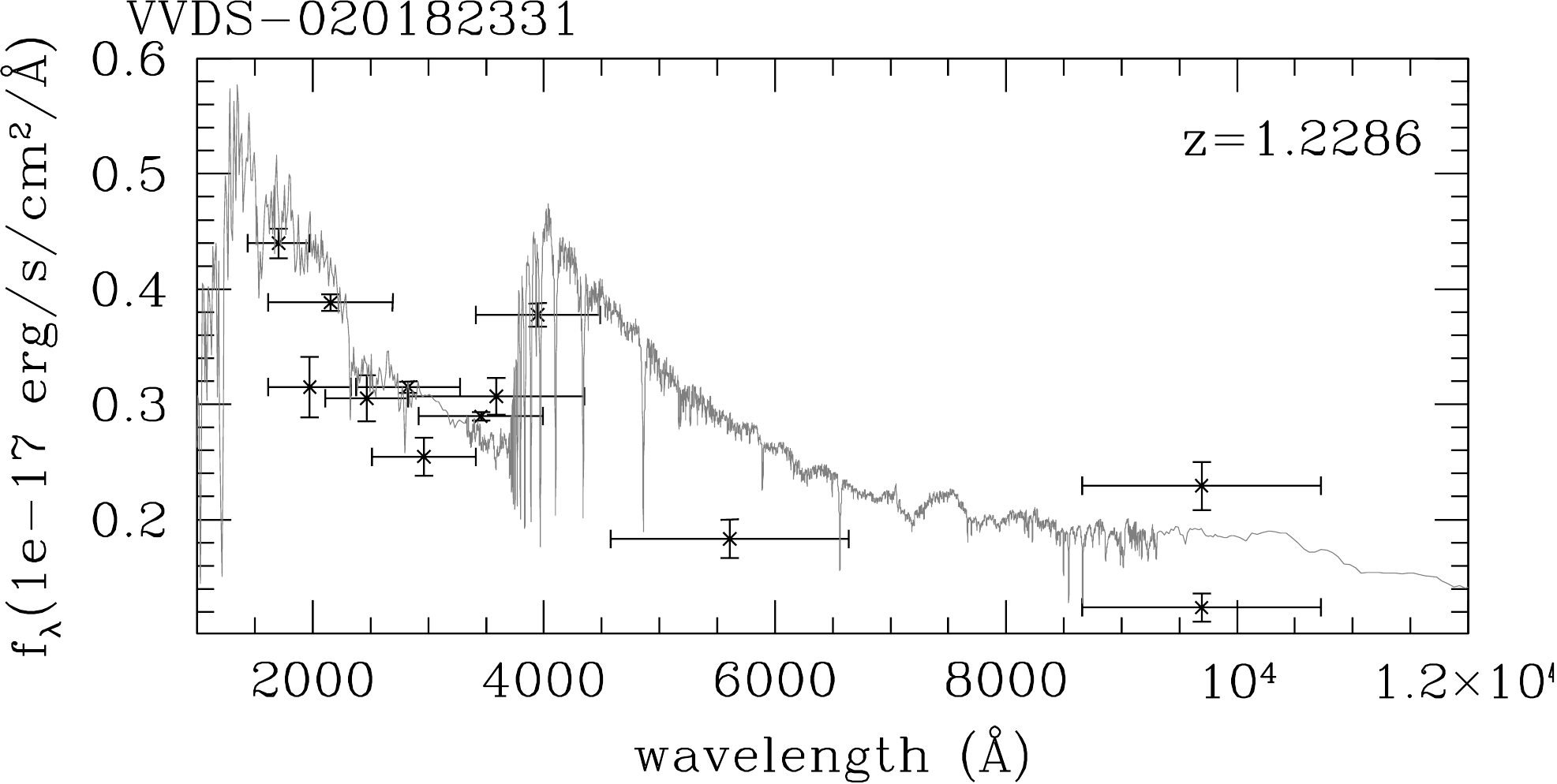}
\includegraphics[width=1\columnwidth]{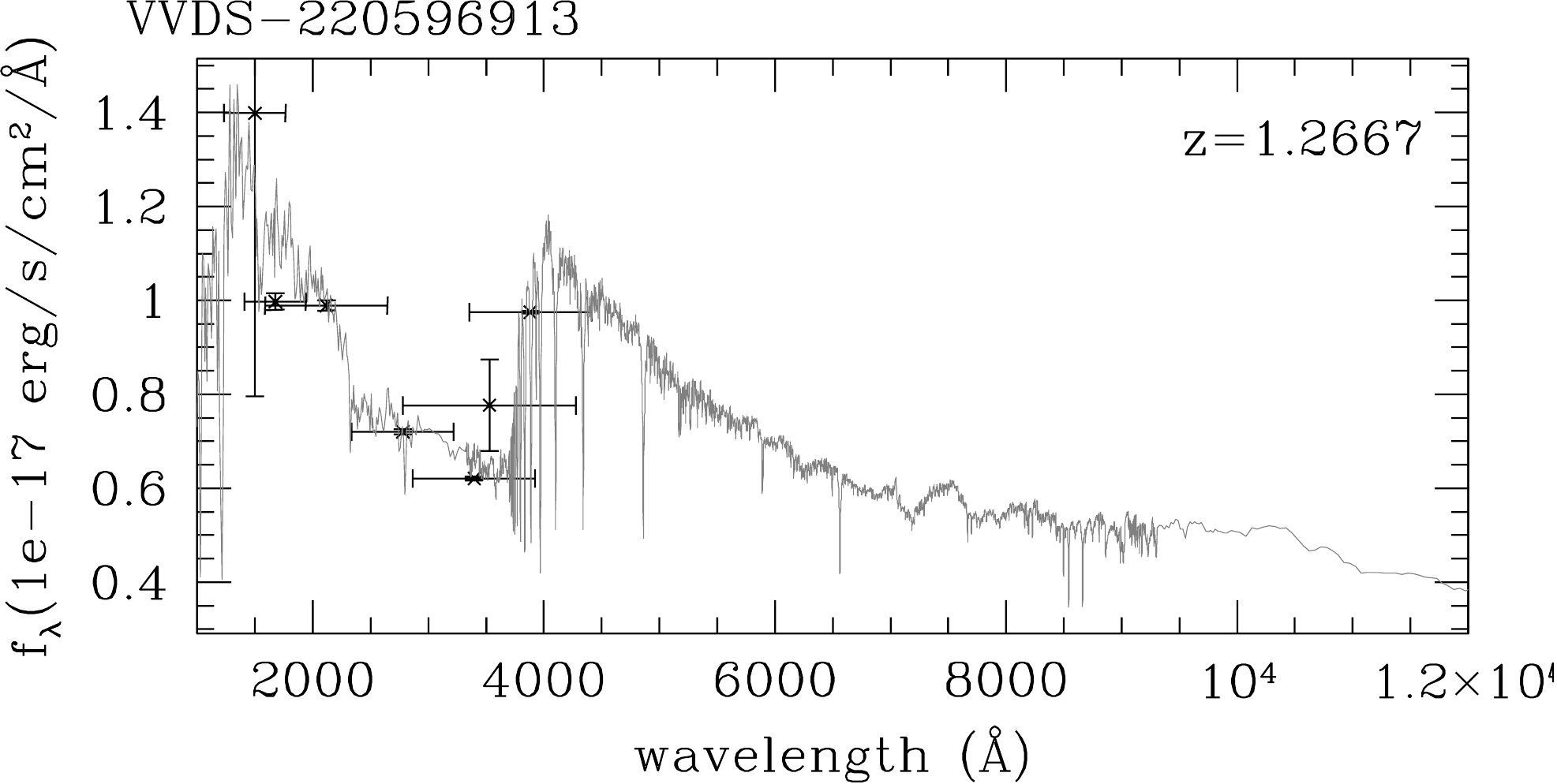} \includegraphics[width=1\columnwidth]{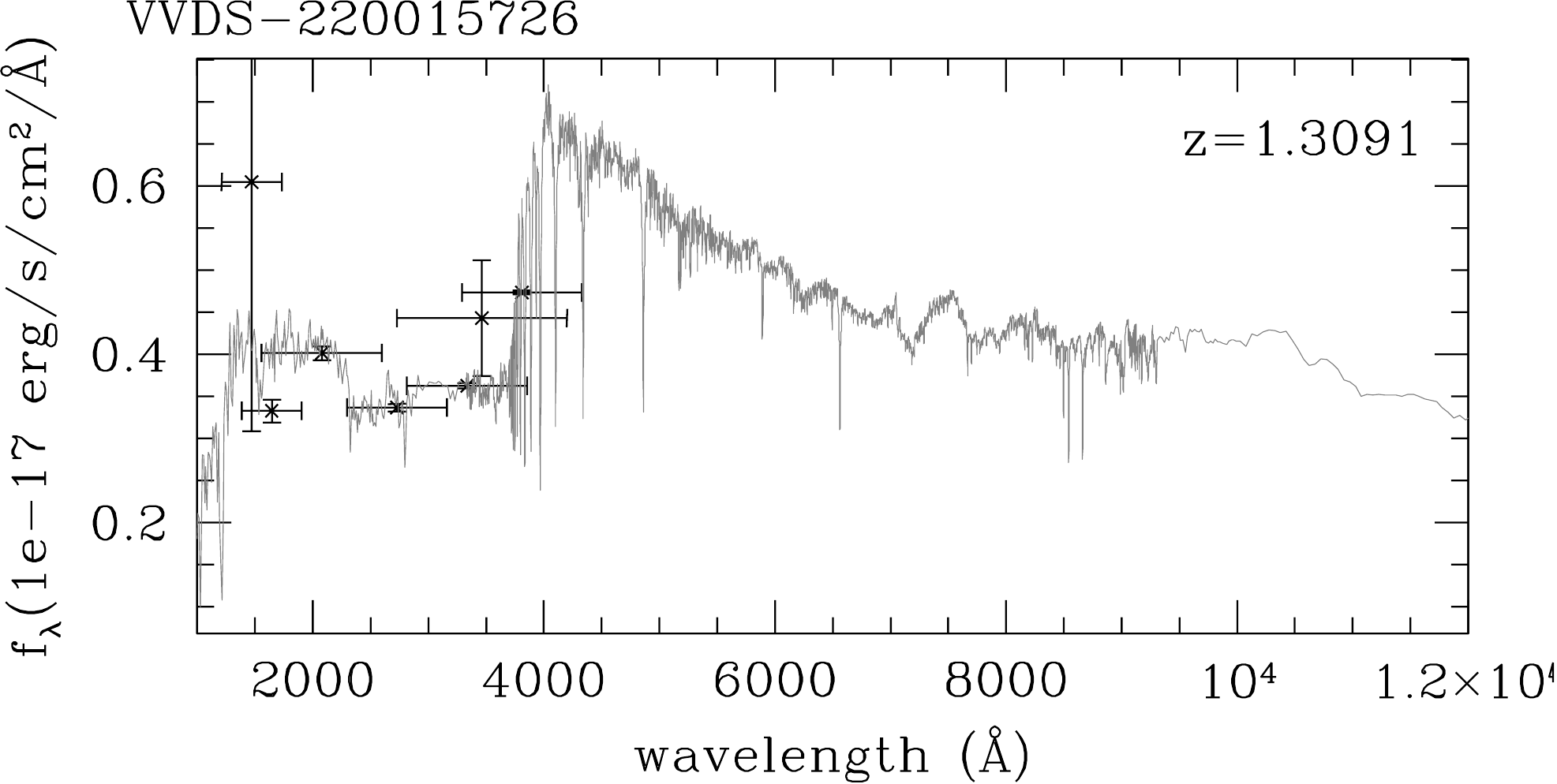}
\includegraphics[width=1\columnwidth]{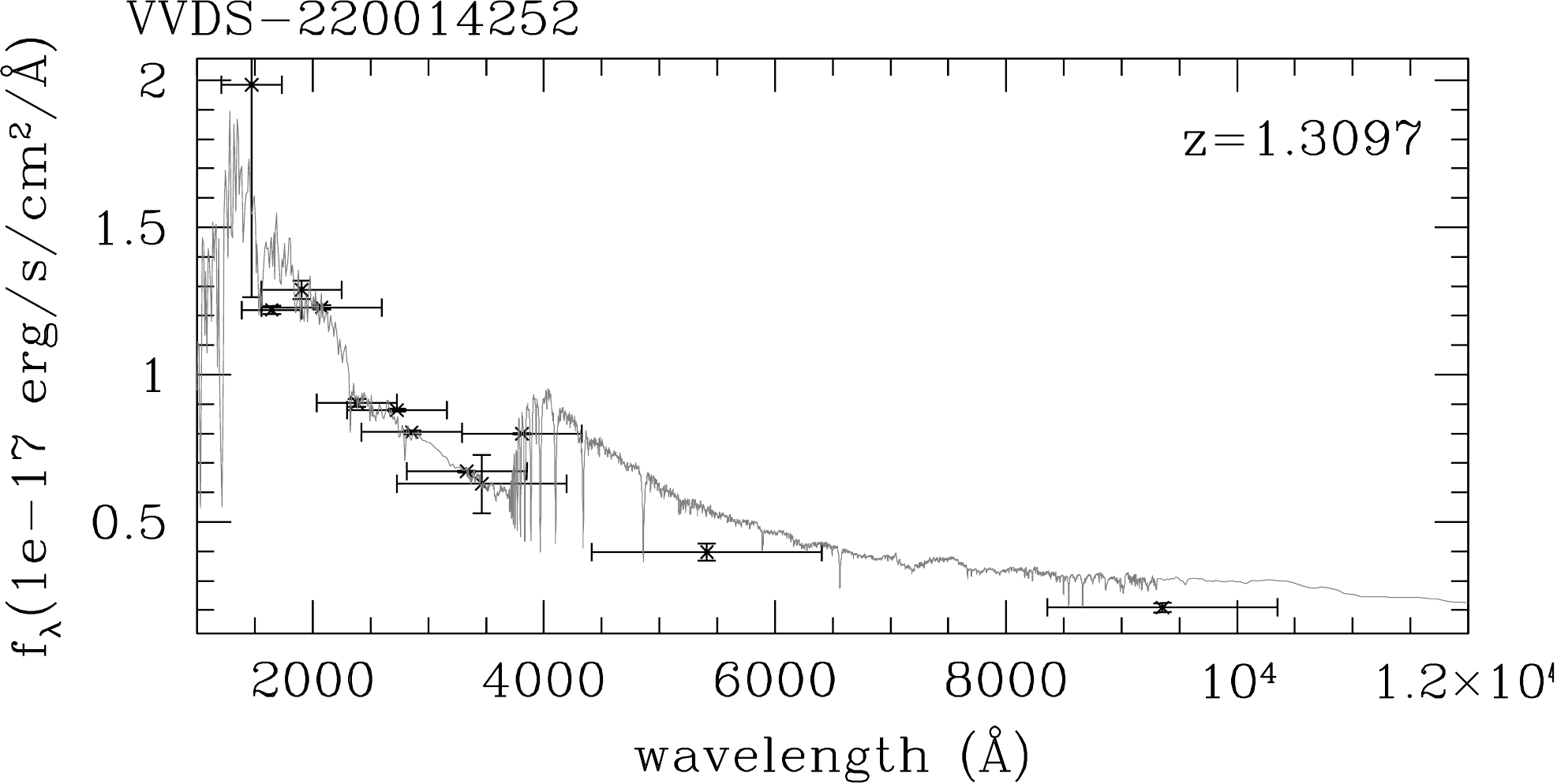} \includegraphics[width=1\columnwidth]{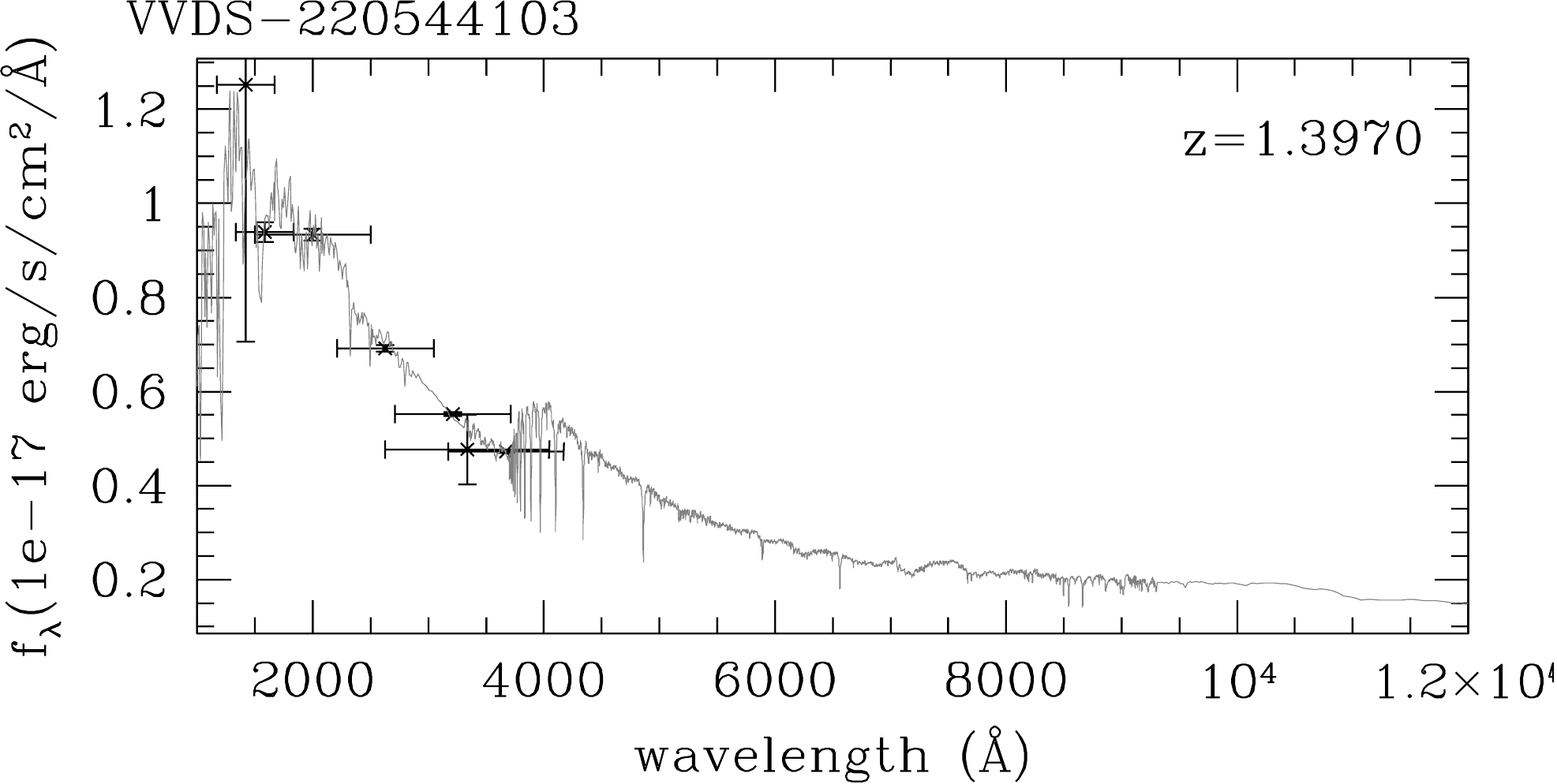}
\includegraphics[width=1\columnwidth]{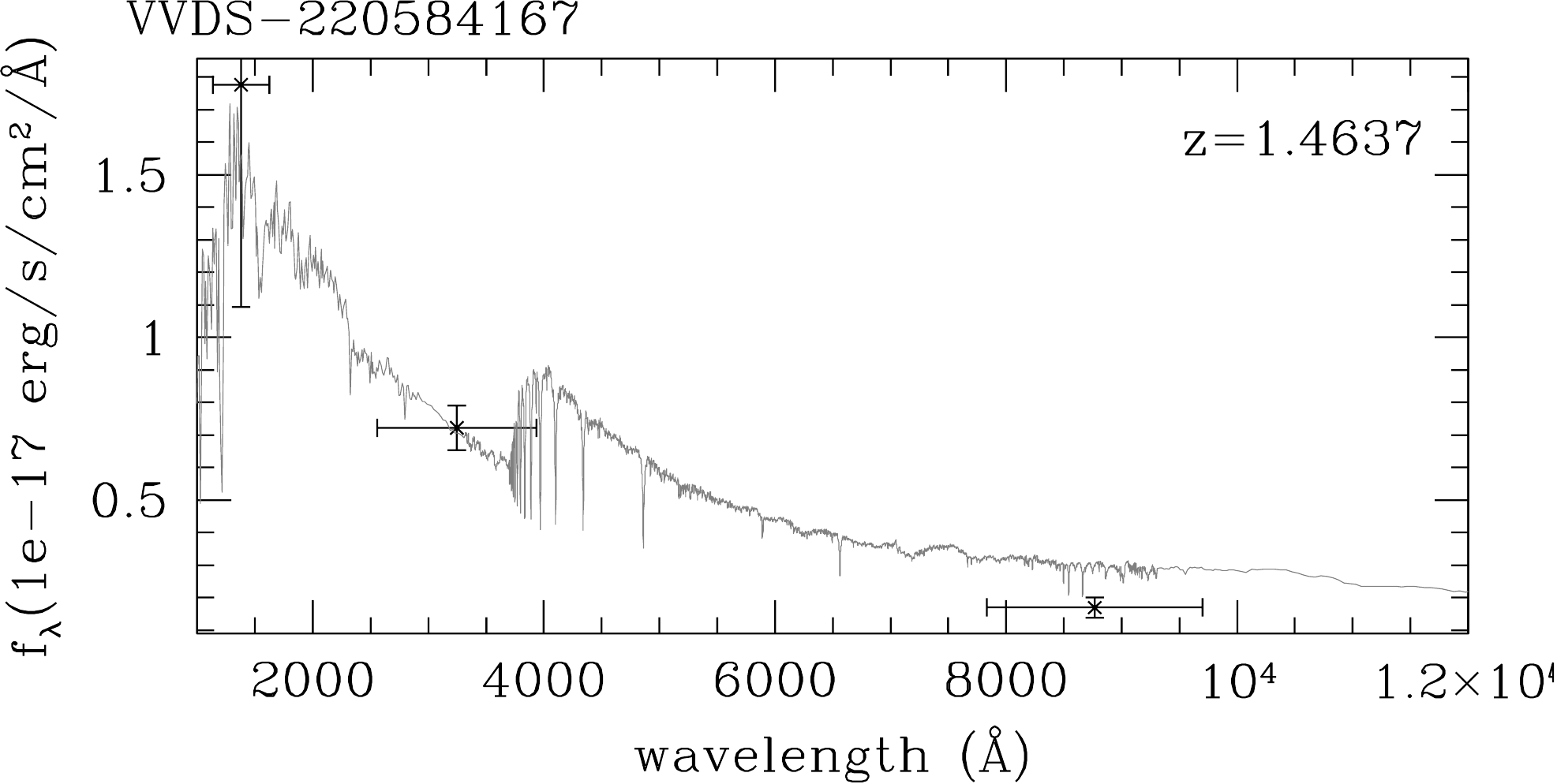} \includegraphics[width=1\columnwidth]{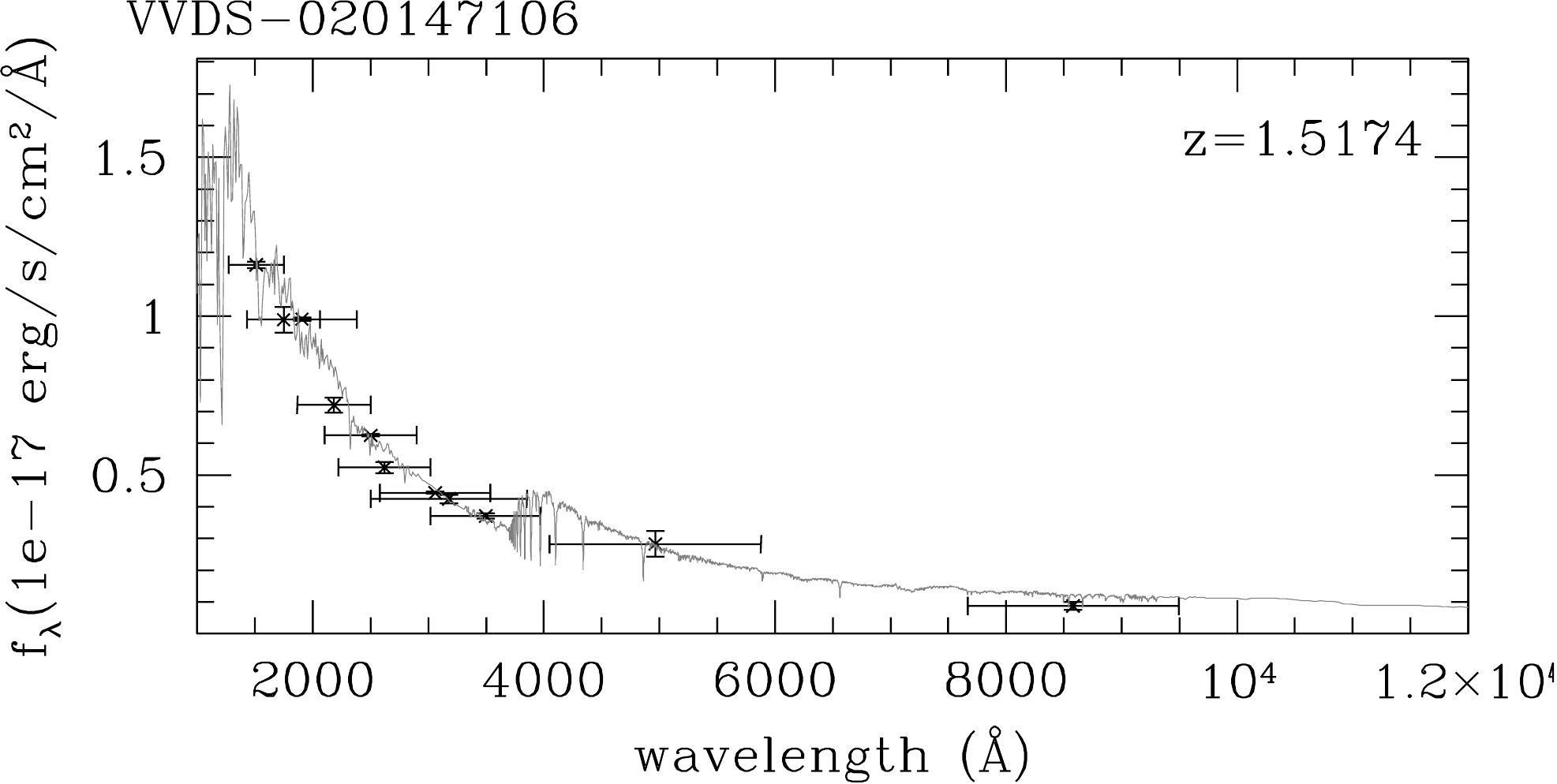}
\includegraphics[width=1\columnwidth]{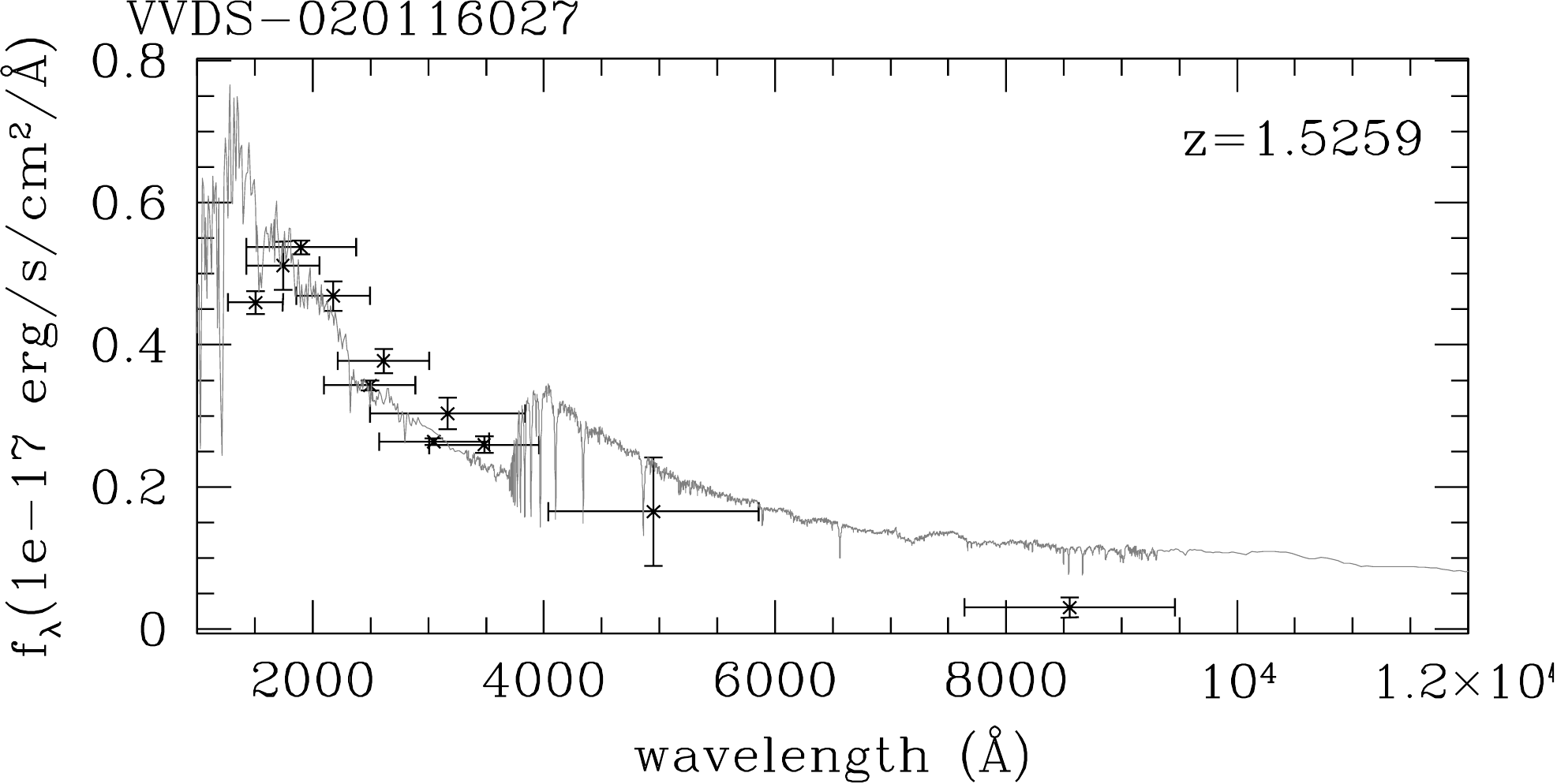} \includegraphics[width=1\columnwidth]{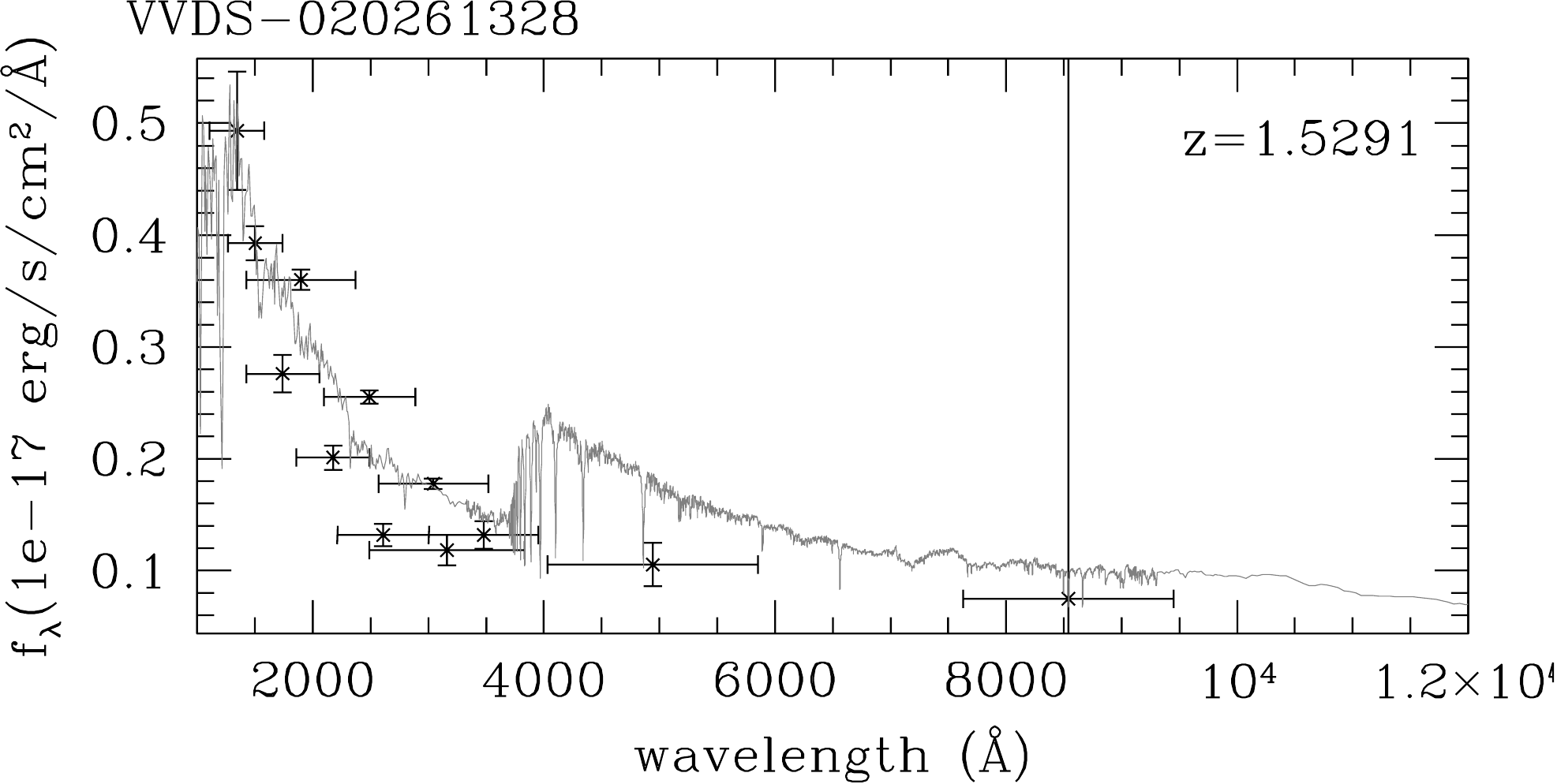} 
\par\end{centering}

\caption{Rest-frame SEDs from the broad-band CFHT optical and near-infrared NTT/SOFI (or UKIDSS) photometry for our sample of 10 galaxies (see also Table~\ref{tabmag}). The best-fitting stellar population model is overlaid.}

\label{figsed}
\end{figure*}

Table~\ref{tabsed} list the properties of the stellar population for the sample of the 10 galaxies derived from the SEDs. The stellar masses, the
reddening suffered by stars, and the ages of the oldest stellar populations are estimated thanks to the full optical and near-infrared photometry
available for our targets. The photometric points are compared to those of a library of synthetic stellar population spectra based on the latest 2007
models of Charlot \& Bruzual (Bruzual 2007, latest models in preparation) models, using the \citet{chabrier:2003PASP..115..763C} IMF. The synthetic
stellar populations are based on exponential declining star formation histories with the addition of random secondary bursts. We calculated the
probability distribution function as e$^{(- \chi ^2)}$ for each observed galaxy compared to all models in the library. We take as an estimate of the
stellar mass the median of this distribution. The results are self-consistently corrected for the effects of age, metallicity and dust. Degeneracies
between these parameters are reflected in the error bars of the derived quantities. Fig.~\ref{figsed} shows the comparison between the observed SEDs and
our best-fitting stellar population models.
%
%
\subsubsection{Star formation rates from photometry}
Ultraviolet-derived star formation rates were calculated from the
 $u$-band photometry (see Table~\ref{tabmag}).
Given the redshift of our targets, the $u$-band is sensitive
to the  rest-frame UV continuum flux
around 1500 \AA ; in the absence of dust, the UV continuum from a young stellar population is approximately flat if the flux is expressed
in frequency units ($f_{\nu}$). 
The mean luminosity in frequency units (\ergshz\ ) is calculated using
the following equation:

\begin{equation}
L_{1500}=10^{(-0.4\cdot m(AB))}\times3631\cdot10^{-23}\times4\pi\frac{D_{L(cm)}^{2}}{1+z}.\end{equation}

We can then deduce $SFR_{UV}$ using the following : 
\begin{equation}
\textup{SFR }_{UV}(M_{\odot}\textup{ yr}^{-1})=0.83\times10^{-28}\; L_{1500}\:\:(\mbox{\ergshz}).\end{equation}
 This equation is the \citet{kennicutt:1998ARA&A..36..189K} calibration
renormalized from \citet{Salpeter:1955ApJ...121..161S} to \citet{chabrier:2003PASP..115..763C}
IMF. Table~\ref{tabsed} list the value of the star formation rate computed from the UV rest-frame continuum emission. 

Additionally the star formation rates, averaged on the last hundred
million years, $SFR_{SED}$, can be also derived from the full SEDs using the same
approach as for the stellar masses, reddening, and ages. They are
also indicated in Table~\ref{tabsed} for the 10 galaxies of our sample.
\subsection{Reddening and dust corrected star formation rate}
The UV star formation rates may be strongly affected by extinction,
its comparison with the H$\alpha$ star formation rate may thus provide
an estimate of the amount of dust. Assuming that all of the ionizing
photons are reprocessed into nebular lines, therefore $SFR_{H\alpha}^{corr}=SFR_{UV}^{corr}$ after correction for dust extinction,
we thus computed the $E(B-V)$ by solving the following system of
two equations: 
\begin{equation}
\begin{array}{ccc}
SFR_{corr} & = & SFR_{H\alpha}\cdot10^{0.4E(B-V)_{gas}\ k^{e}(0.6563)}\\
SFR_{corr} & = & SFR_{UV}\cdot10^{0.4E(B-V)_{gas}\ k^{e}(0.1500)}\end{array},
\end{equation}
 where the obscuration curve for the ionized gas, k$^{e}$($\lambda$),
is given by \citet{Calzetti:2001PASP..113.1449C}:

\begin{equation}
\begin{array}{ll}
k^{e}(\lambda)= & 2.659\,(-2.156+1.509/\lambda-0.198/\lambda^{2}\\
 & +0.011/\lambda^{3})+4.05\\
 & \,\,\,\,\,\,\,\,\,\,\,\,\,\,\,0.12\ \mu m\le\lambda<0.63\ \mu m,\\
k^{e}(\lambda)= & 2.659\,(-1.857+1.040/\lambda)+4.05\\
 & \,\,\,\,\,\,\,\,\,\,\,\,\,\,\,0.63\ \mu m\le\lambda\le2.20\ \mu m.
\end{array}\label{calz}
\end{equation}
We, then, derived the deredenned star formation rate, $SFR_{corr}$, listed in Table~\ref{tabline}.
%
\subsection{Gas mass and gas mass fraction}
The relation between the star formation surface density $\Sigma_{SFR}$
(in $M_{\odot}\,\mbox{yr}^{-1}\,\mbox{kpc}^{-2}$) and the gas surface
density $\Sigma_{gas}$ (in $M_{\odot}\,\mbox{pc}^{-2}$) is given
by the Schmidt law \citep{kennicutt:1998ARA&A..36..189K}: \begin{equation}
\Sigma_{SFR}=1.48\cdot10^{-4}\times(\Sigma_{gas})^{1.4},\label{eq:schmidt}\end{equation}
which has been renormalized to the \citet{chabrier:2003PASP..115..763C}
IMF. We thus calculate the star formation surface density by dividing
the deredenned star formation rate ($SFR_{corr}$) by the projected area of the galaxy
(see Table~\ref{tabmorpho}). We used the two-dimensional
\halpha\ emission distribution, deconvolved with the point spread function (PSF), to directly
estimate the spatial extend of the ionized gas, $A_{gas}$, so that
$M_{gas}=A_{gas}\times\Sigma_{gas}$.

We then calculate the gas surface density by inverting the above equation,
and finally calculate the gas mass by multiplying the gas surface
density with the projected area. We also derived the gas mass fraction,
which is $\mu=M_{gas}/(M_{gas}+M_{\star})$, where $M_{\ast}$ is
the stellar masses deduced from the SED modelling. The gas mass and
the gass mass fraction are given in Table~\ref{tabline}. We found
that six of our galaxies have high values of the gas fraction ($\mu>0.5$).
%
\subsection{Virial mass}
While the global velocity dispersion $\sigma_{1D}$ alone does not allow us
to distinguish between any systematic velocity shift across the galaxy
and any random motions, it can provide a reasonable estimate of the
mass of the object. Assuming a virialized motion around the morphological
galaxy centre, we can use this $\sigma_{1D}$ as a crude estimate
of the mass within the largest radius $r_{gas}$ (estimated from the 2$\sigma$ width
of the H$\alpha$ flux profile along the major axis and deconvolved with the seeing, see Table~\ref{tabmorpho})
for which a velocity is measured using the formula: \begin{equation}
M_{vir}(\sigma_{1D})=\frac{C\sigma_{{\rm 1D}}^{2}r_{gas}}{G}\label{dynmass.eqn}\end{equation}
 (where $C=3.4$ for of a gas-rich disk with an average inclination
angle; see  \citet{erb:2006ApJ...647..128E}). In
Table~\ref{tabkine} we indicate dynamical masses inferred from the
$\sigma_{1D}$. The values span a range from $\sim$1-26x10$^{10}\, M_{\odot}$, with four galaxies having a virial mass $>1$x10$^{11}\, M_{\odot}$
\begin{table*}
\caption{Morphological parameters and initial guesses for the rotation modelling.}
\begin{tabular}{ccccccc}
\hline 
Galaxy  & $\Delta c$ (a)  & PA (b)  & $\Delta$PA$_{gas}$ (c)  & $i$ (d)  & $r_{gas}$ (e)  & $A_{gas}$ (f)\tabularnewline
\hline 
VVDS-1235  & -  & $42\lyxmathsym{\textdegree}$$^{\star}$  & -  & $70\lyxmathsym{\textdegree}$  & $9.8\pm1.4$  & $63$\tabularnewline
VVDS-2331  & $2.4$  & $-69\lyxmathsym{\textdegree}$  & $-41\lyxmathsym{\textdegree}$  & $52\lyxmathsym{\textdegree}$  & $9.2\pm1.6$$^{\dagger}$  & $120$\tabularnewline
VVDS-6913  & $8.5$  & $-99\lyxmathsym{\textdegree}$  & $-4\lyxmathsym{\textdegree}$  & $63\lyxmathsym{\textdegree}$  & $20.2\pm0.8$  & $348$\tabularnewline
VVDS-5726  & $1.7$  & $-18\lyxmathsym{\textdegree}$  & $-10\lyxmathsym{\textdegree}$  & $24\lyxmathsym{\textdegree}$  & $6.7\pm1.0$  & $112$\tabularnewline
VVDS-4252  & $1.8$  & $45\lyxmathsym{\textdegree}$  & $-3\lyxmathsym{\textdegree}$  & $50\lyxmathsym{\textdegree}$  & $10.6\pm1.1$  & $225$\tabularnewline
VVDS-4103  & $2.6$  & $-167\lyxmathsym{\textdegree}$  & $-18\lyxmathsym{\textdegree}$  & $53\lyxmathsym{\textdegree}$  & $13.4\pm1.2$  & $315$\tabularnewline
VVDS-4167  & $2.8$  & $162\lyxmathsym{\textdegree}$  & $-10\lyxmathsym{\textdegree}$  & $62\lyxmathsym{\textdegree}$  & $16.6\pm1.4$  & $468$\tabularnewline
VVDS-7106  & $0.0$ & $-119\lyxmathsym{\textdegree}$ & $0\lyxmathsym{\textdegree}$ & $37\lyxmathsym{\textdegree}$ & $7.7\pm1.6$ & $142$\tabularnewline
VVDS-6027  & -  & $175\lyxmathsym{\textdegree}$$^{\star}$  & -  & $41\lyxmathsym{\textdegree}$  & $6.9\pm1.1$  & $108$\tabularnewline
VVDS-1328  & $1.2$  & $-27\lyxmathsym{\textdegree}$  & $30\lyxmathsym{\textdegree}$  & $20\lyxmathsym{\textdegree}$  & $7.1\pm1.1$  & $127$\tabularnewline
\hline
\end{tabular}

\begin{raggedright}
The columns are as follows: (a) distance between the maximum of \halpha\ distribution
and the kinematic centre (kpc); (b) initial guess kinematic position
angle; (c) difference between the morphological (inferred from the
\halpha\ distribution) and kinematic position angles; (d) inclination
inferred from the \halpha\ distribution ($0\lyxmathsym{\textdegree}$
for face-on); (e) 2$sigma$ profile of \halpha\ distribution map
along the major axis (kpc). Uncertainty represents half
of the PSF correction; (f) projected area of the nebular emission
(kpc$^{2}$). \\
 $^{\star}$: for objects which do not show a rotation, we give
the morphological PA. \\
 $^{\dagger}$: for VVDS-2331, we give the radius calculated
with respect to the morphological major axis, not the kinematical
major axis.
\par\end{raggedright}

\label{tabmorpho}
\end{table*}
%
\begin{table*}
\caption{Best-fit parameters from the rotation modelling.}
\begin{tabular}{ccccccccc}
\hline 
Galaxy  & PA$_{mod}$ (a)  & $i_{mod}$ (b)  & $\sigma_{0}$ (c)  & $V_{s}$ (d)  & $V_{rot}$ (e)  & $r_{c}$ (f)  & res(v) (g)  & res(d) (h)\tabularnewline
\hline 
VVDS-2331  & $-64\lyxmathsym{\textdegree}\pm21$  & $51\lyxmathsym{\textdegree}\pm33$  & $38\pm41$  & $-8\pm21$  & $193\pm72$  & $4.8\pm8$  & $-1\pm43$  & $11\pm34$\tabularnewline
VVDS-6913  & $-113\lyxmathsym{\textdegree}\pm28$  & $62\lyxmathsym{\textdegree}\pm19$  & $44\pm22$  & $-6\pm19$  & $139\pm30$  & $10.7\pm3$  & $-2\pm30$  & $20\pm23$\tabularnewline
VVDS-5726  & $-6\lyxmathsym{\textdegree}\pm22$  & $24\lyxmathsym{\textdegree}\pm7$  & $60\pm35$  & $-17\pm15$  & $292\pm71$  & $1.5\pm2$  & $-2\pm25$  & $5\pm19$\tabularnewline
VVDS-4252  & $32\lyxmathsym{\textdegree}\pm11$  & $48\lyxmathsym{\textdegree}\pm15$  & $100\pm12$  & $-2\pm7$  & $130\pm23$  & $4.5\pm2$  & $4\pm10$  & $0\pm11$\tabularnewline
VVDS-4103  & $167\lyxmathsym{\textdegree}\pm23$  & $79\lyxmathsym{\textdegree}\pm5$  & $58\pm37$  & $-10\pm12$  & $118\pm27$  & $8.5\pm3$  & $-2\pm20$  & $1\pm28$\tabularnewline
VVDS-4167  & $-176\lyxmathsym{\textdegree}\pm16$  & $72\lyxmathsym{\textdegree}\pm23$  & $54\pm25$  & $-11\pm12$  & $257\pm34$  & $10.8\pm2$  & $2\pm14$  & $5\pm18$\tabularnewline
VVDS-7106  & $-134\lyxmathsym{\textdegree}\pm35$  & $37\lyxmathsym{\textdegree}\pm26$  & $82\pm6$  & $7\pm6$  & $105\pm71$  & $9.4\pm4$  & $2\pm7$  & $-2\pm10$\tabularnewline
VVDS-1328  & $-10\lyxmathsym{\textdegree}\pm25$  & $20\lyxmathsym{\textdegree}\pm7$  & $36\pm17$  & $17\pm10$  & $182\pm56$  & $0.5\pm4$  & $-1\pm11$  & $7\pm17$\tabularnewline
\hline
\end{tabular}

The columns are as follows: (a) position angle (b) inclination (c)
local velocity dispersion (km s$^{-1}$) (d) systemic velocity (km s$^{-1}$) (e) maximum
velocity (km s$^{-1}$) (f) radius where the maximum velocity is reached (kpc)
(g) mean value and rms of the residual velocity map (km s$^{-1}$) (h) mean
value and rms of the residual dispersion map (km s$^{-1}$).
\label{tabmodel}
\end{table*}
\section{Properties from spatially resolved kinematics} 
\label{kine_prop} 
\subsection{H$\alpha$ kinematics}
\begin{figure*}
\begin{centering}
\resizebox{2.0\columnwidth}{!}{\includegraphics{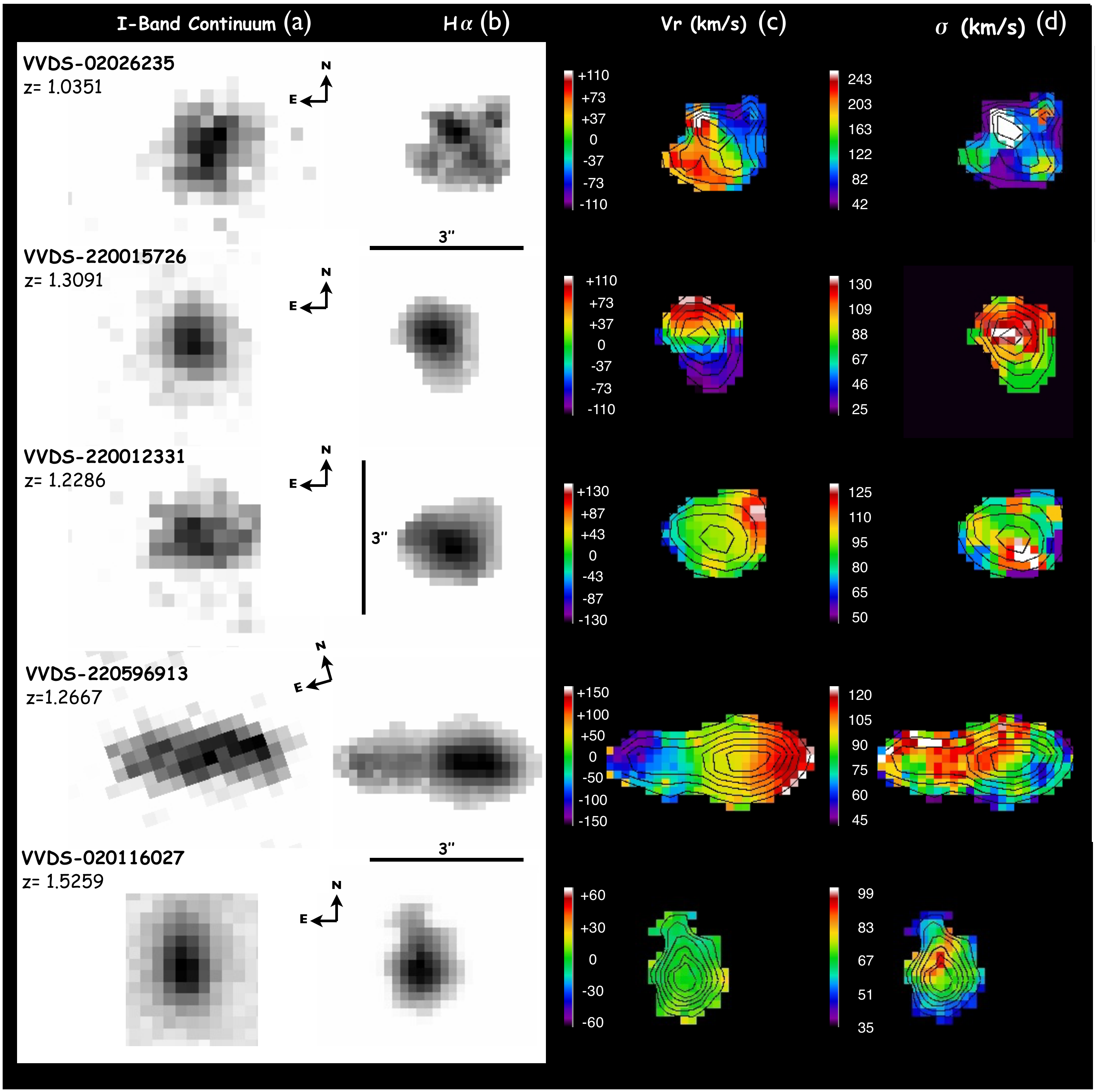}} 
\par\end{centering}

\caption{From left to right: (a) $I$-band CFHT image, (b) H$\alpha$
flux map, (c) H$\alpha$ velocity field, and (d) H$\alpha$ velocity
dispersion map obtained from Gaussian fits to the SINFONI data cubes
after smoothing spatially with a two-dimensional Gaussian of FWHM
$=$ 3 pixels. The $I$-band image and H$\alpha$ maps are color-coded
with a linear scaling such that the values increase from light to
dark. These data have been acquired with the 125$\times$250mas sampling
configuration of SINFONI, in seeing limited mode. An angular size
of 1\arcsec corresponds to $\sim8$ kpc at the redshifts of most of the
objects. The direction of North-East, the VVDS number identification and
VIMOS-based redshift are indicated for each galaxy.} 
\label{map1_fig}
\end{figure*}
We produced two-dimensional maps of the dynamics for the galaxies
in our sample by using E3D, the Euro3D visualization tool \citep{Sanchez:2004AN....325..171S},
and a code for the fitting and analysis of kinematics \citep[e.g. ][]{Sanchez:2004ApJ...615..156S,Sanchez:2005A&A...429L..21S}.
Our main goal was to determine the kinematics of the ionised gas using
the strongest emission line, \halpha . For each spatial pixel (spaxel)
we fit the \halpha\ emission line profile with a single Gaussian function,
in order to characterize the emission line, and a pedestal to characterise
any spectral continuum. We first smoothed the reduced cubes spatially
with a Gaussian of ${\rm FWHM=3~pixels}$ ($0\farcs37$). The line
flux, FWHM, central wavelength and the continuum (pedestal) were then
fitted. From the results of this fitting we obtained maps (see Fig.~\ref{map1_fig}
and Fig.~\ref{map2_fig}) of the \halpha\ emission line intensity, the
relative radial velocity map ($V_{r}$) and the velocity dispersion
($\sigma$). The dispersion map was corrected for the contribution
of the instrumental dispersion, as determined from the FWHM of unblended
and unresolved sky lines. Error maps for the velocity and dispersion
measurements were also computed. The errors are dominated by the effect
of random noise in fitting the line profile which produces large
errors at low S/N. Errors ranged from $~4$ to $~25$
km s$^{-1}$ for both the radial velocity and velocity dispersion maps.
%
\subsection{Gas morphology}
\label{morpho_prop}
\begin{figure*}
\begin{centering}
\resizebox{2.0\columnwidth}{!}{\includegraphics{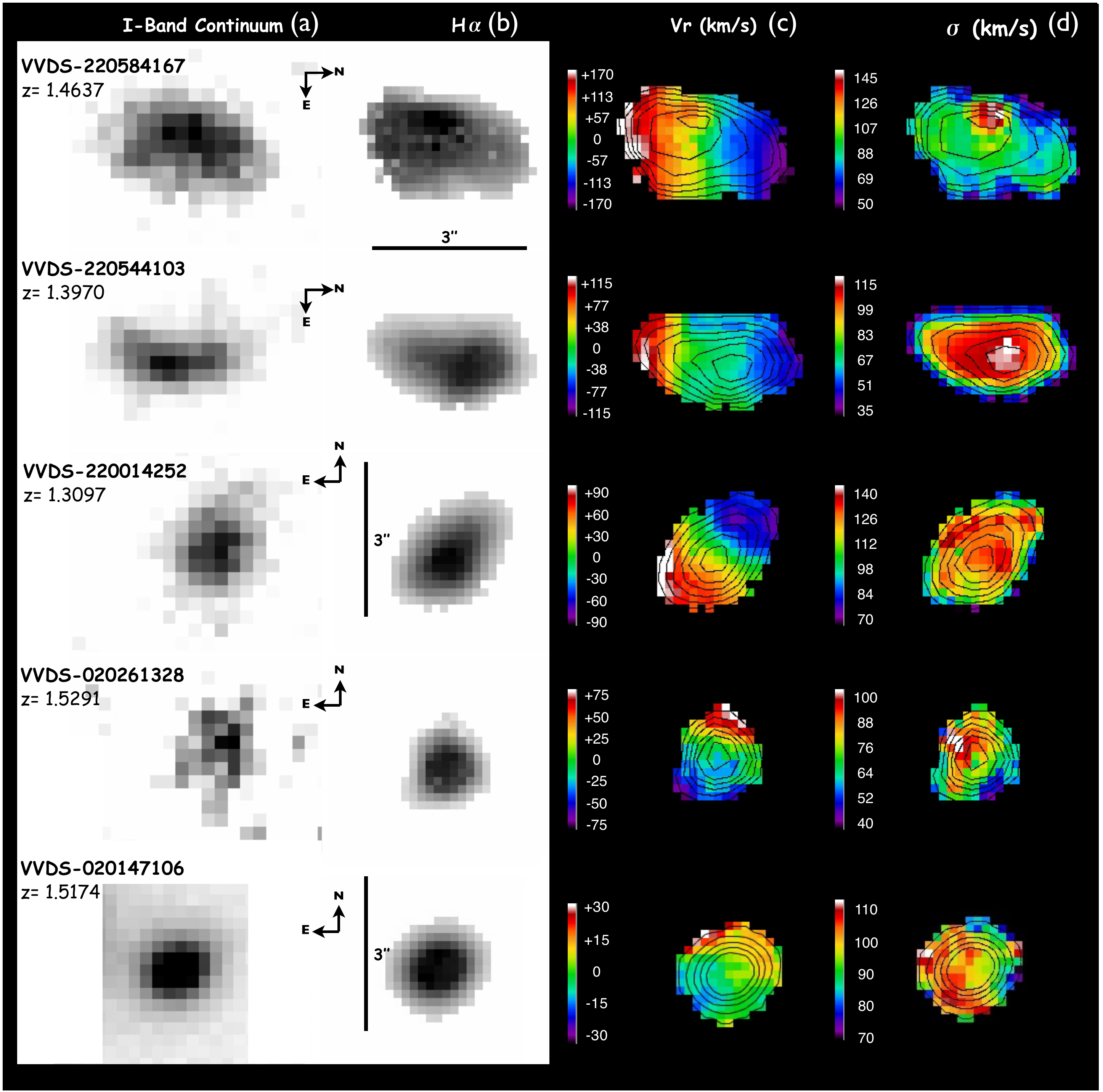}} 
\par\end{centering}

\caption{Similar as Fig.\ref{map1_fig} for five other galaxies.}
\label{map2_fig}
\end{figure*}

Fig.~\ref{map1_fig} and Fig.~\ref{map2_fig} contain images of the \halpha\ emission
line intensity for each of the galaxies. We also plot 
the morphology of the stellar component mapping by the CFTHLS $I$-band
continuum. Nine of the objects have a distinct single peak in the
\halpha\ distribution. This peak emission of the star formation
from \halpha\ is coincident (or near-coincident)
with the peak stellar continuum emission in the $I$ band. Only one of
the targets, VVDS-1235, has distinct resolved knot-like
structures in the \halpha\ distribution which are not seen in the
stellar emission.

Given the similarity between the distribution of the H$\alpha$ flux
in our IFU observations and the continuum in the optical images, we
decided to use our line flux maps, which have the better resolution,  to infer the
morphological parameters: the centre (as defined by the position where
the flux is maximum); the position angle (PA), and the inclination. The
inclination is calculated as $\cos^{-1}(b/a)$, where $a$ and \textbf{$b$}
are the radius of major and minor axis, respectively, which are measured
as the 1$\sigma$ profile of the flux map. 
The distance
between the kinematical and the morphological centres is given in Table~\ref{tabmorpho}.
We also measure a kinematical PA in case it is slightly
offset from the morphological PA. The kinematical PA can be measured in two consistent ways: or we look for the major axis
as the direction where the velocity gradient is maximum, or we look
for the minor axis as the direction where the velocity gradient is
minimum. The difference between the kinematical and morphological
PAs is given in Table~\ref{tabmorpho}. 
%
\subsection{Rotation modelling}

Each galaxy is modelled by a pure, infinitely thin, rotating disk. The parameters of the model are the kinematic centre ($x_{0}$,$y_{0}$), the
position angle (PA), the inclination ($i$), the velocity offset ($V_{\mathrm{s}}$) of the centre relative to the integrated spectrum, and the
velocity curve $V_{\mathrm{c}}(r)$ where $r$ is the radius from the kinematic centre. We also use the true physical velocity dispersion
$\sigma_{0}$ as a model parameter, which we assume is constant, and represents the thickness of the rotating disk.
Below its maximum, the velocity is assumed to increase linearly with
the radius and then remain constant \citep{Wright:2007ApJ...658...78W}.
 The radial velocity $V$ for
any point is then defined with standard projection equations. Note that the PA gives the direction of positive radial velocities.
The velocity offset accounts for redshift uncertainties. We define the velocity curve by two parameters, following \citet{Wright:2007ApJ...658...78W}: the maximum velocity $V_{\mathrm{max}}$ and the radius $r_{\mathrm{c}}$ where this maximum velocity is achieved. In this model, below
$r_{\mathrm{c}}$ the velocity is defined as:
\begin{equation}
V_{\mathrm{c}}(r) = \frac{r}{r_{\mathrm{c}}}\times V_{\mathrm{max}}\
\end{equation}
The model computes
$V_{rot}$, the asymptotic maximum rotation velocity at the plateau of the rotation curve, corrected for inclination ($V_{rot}=V_{rot}/\sin i$).

In reality, the spatial resolution is limited by the seeing and the spaxel size. The observed radial velocity is thus the weighted convolution of the
true radial velocity by the PSF. This PSF is modelled as a 2D Gaussian and takes into account the 3-pixel spatial smoothing. The weights come from the
flux map of the line used to compute the velocity map: a spaxel where the observed line flux is negligible will not contribute to the convolution.
Additionally, the observed velocity dispersion accounts not only for the true physical dispersion, but also for the variations in the velocity field
inside the width of the PSF. We generate a map of the velocity dispersion where the velocity gradient (determined from the modelling) has been
subtracted. This yields a better measure of the intrinsic local velocity dispersion, rather than a raw value of $\sigma$ which may be inflated if the
velocity gradient across a spaxel is large. From this map, we compute the flux-weighted mean velocity dispersion, $\sigma_{o}$ (Table~\ref{tabmodel}).

The beam smearing introduced by the PSF causes two significant effects. First, the observed velocity map appears smoothed, therefore the velocity
gradient and the maximum velocity are underestimated. Secondly, the observed dispersion map shows a peak of dispersion near to the kinematic centre. We
can reproduce modelled velocity and dispersion maps by applying mathematically the same weighted convolution to the ideal velocity field. A better
modelling of the beam smearing would be to convolve our modelled velocity and dispersion maps with the PSF in each wavelength channel, weighted by the
intrinsic flux distributions of the systems at this wavelength \citep{1994A&A...285..723E}. However, for such high redshift galaxies for which the
intrinsic flux distributions is unknown and taking into account the quality of the data, it is not possible to model the velocity and dispersion maps to
such a high precision. The modelled maps are compared to the observed maps by a $\chi^{2}-$minimization. The parameters which minimize the $\chi^{2}$
are computed in two successive grids with increasing resolution. First guesses for the kinematic centre, the PA, and the inclination are set from the
morphology.
\begin{figure}
\begin{centering}
\includegraphics[width=1\columnwidth]{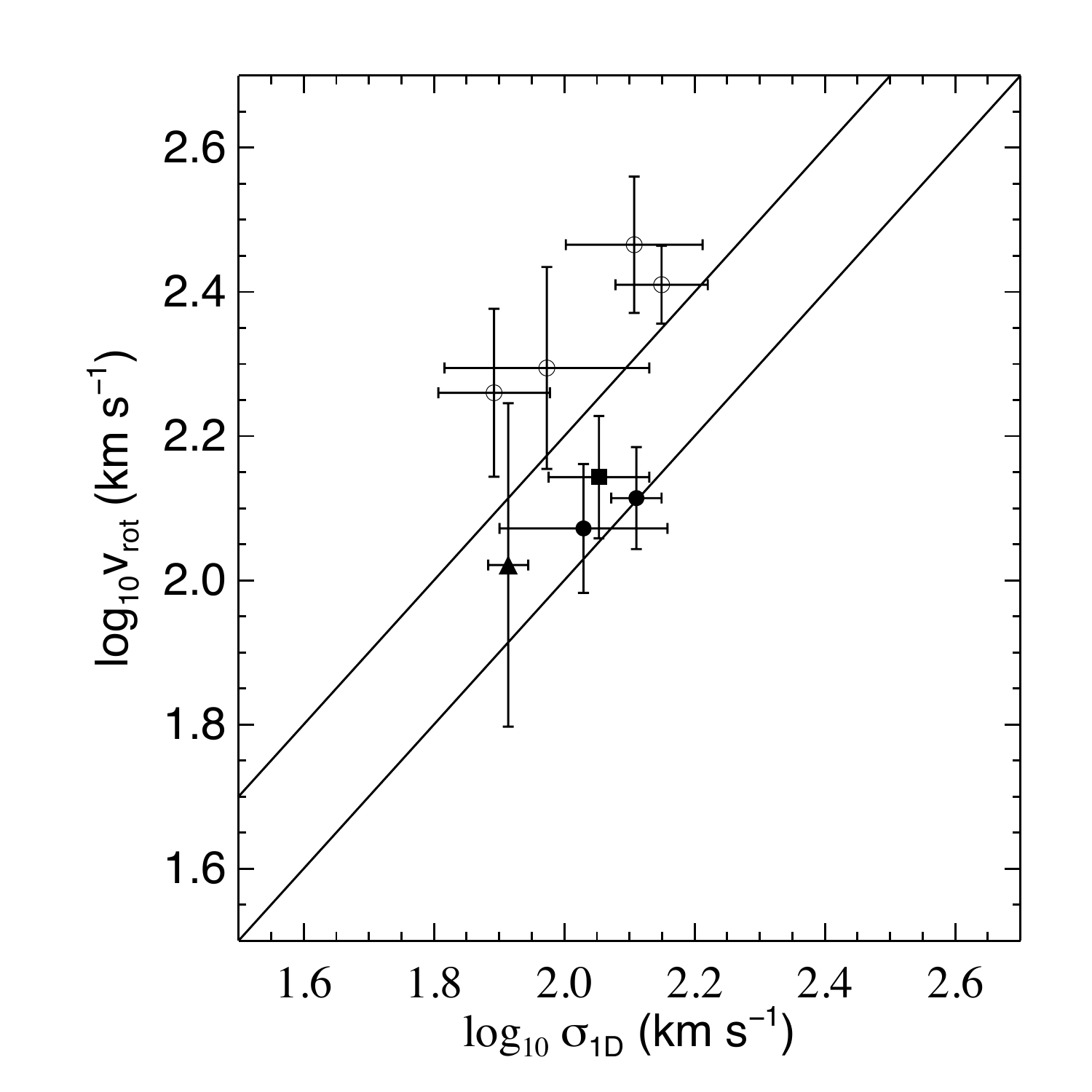}
\par\end{centering}
\caption{Maximum velocity from rotation modeling (log $V_{rot}$) vs integrated  line width (log $\sigma_{1D}$). \emph{open circle}: RD rotating disks and \emph{filled symbols}: DD rotating disks. Among the DD rotating disks we highlighted in \emph{filled triangle}: VVDS-7106, and in  \emph{filled square}: VVDS-6913. The lines are the 1:1 line and the Rix et al. (1997) $\sigma = 0.6 V_{c}$ line.}
\label{Fig:weiner16}
\end{figure}
\begin{figure*}
 \begin{centering}
\resizebox{2.1\columnwidth}{!}{\includegraphics{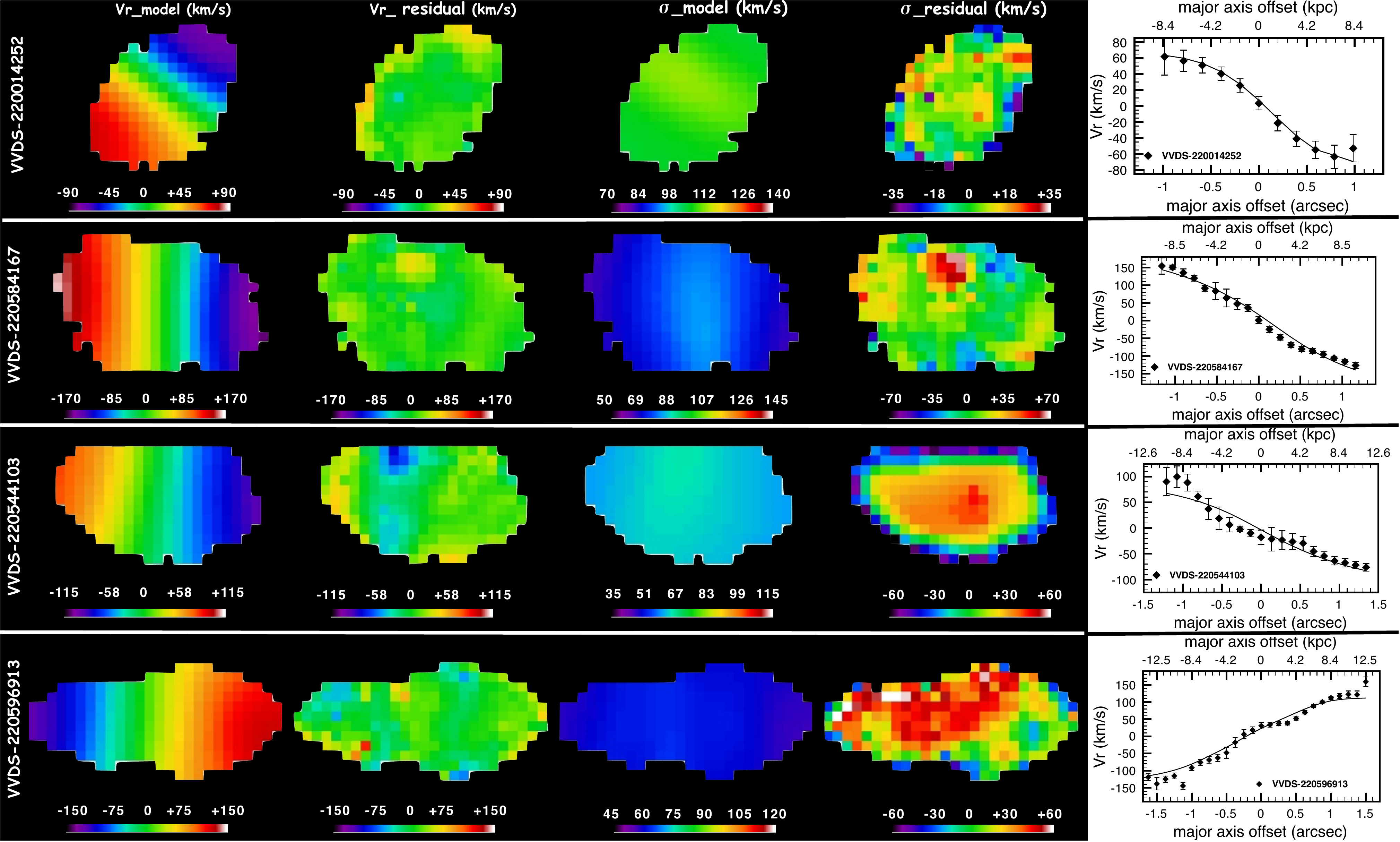}} 
\par\end{centering}
\caption{Kinematic best-fit models of galaxies of our sample (left panel) for both the velocity and the dispersion, after adding the effect of the beam smearing.The difference for both the velocity and the dispersion between the observed map and the best-fit model map is also shown and may display non-negligible random motions in the gas.   An one-dimensional rotation curve extracted using an `idealized' slit from the observed \halpha\ velocity map, overlaid with the best-fitting model, is shown in the right panel.}
\label{map3_mod}
\end{figure*}
\begin{figure*}
\begin{centering}
\resizebox{1.5\columnwidth}{!}{\includegraphics{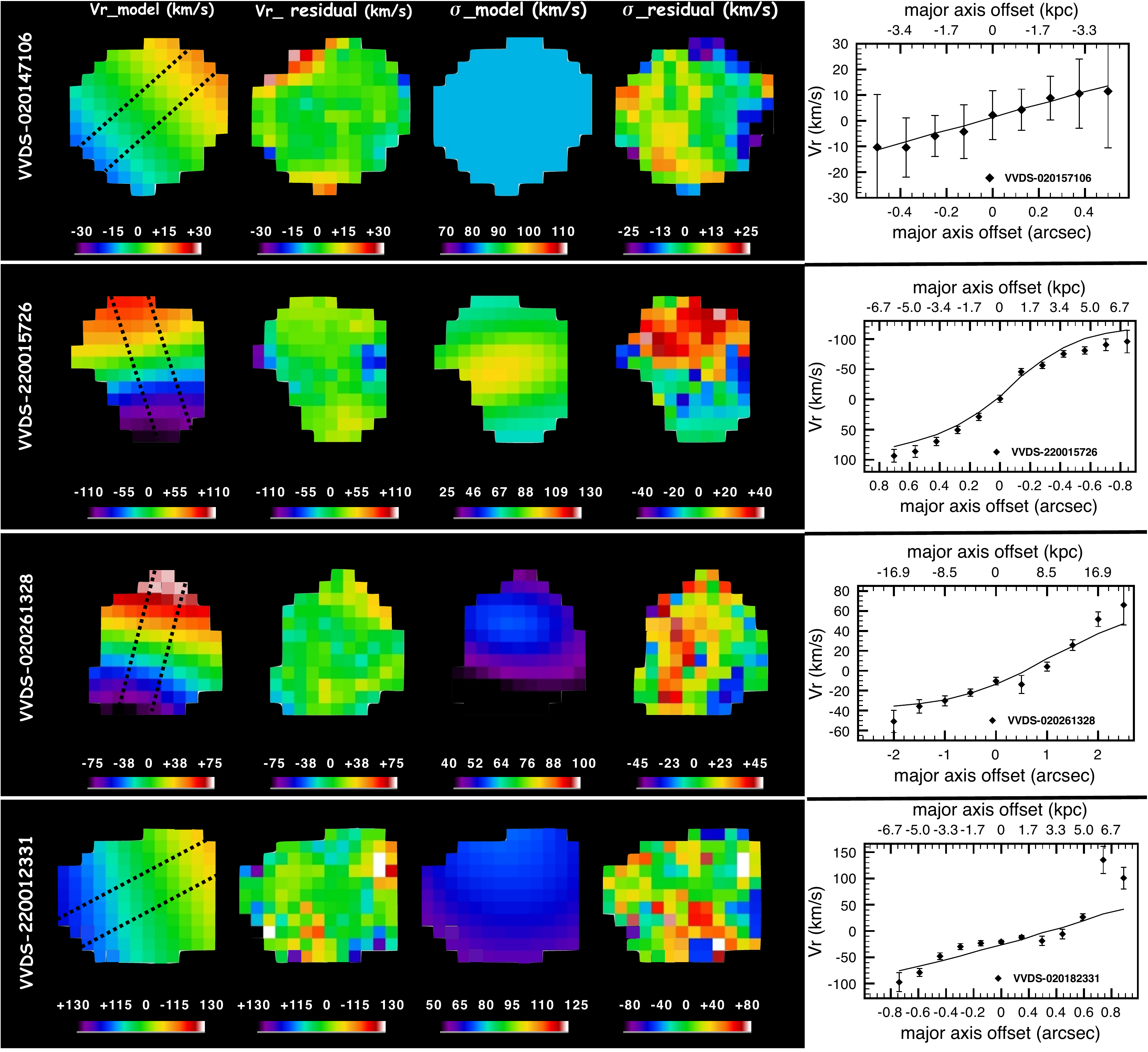}} 
\par\end{centering}
\caption{Similar as Fig.\ref{map3_mod} for four other galaxies.}
\label{map4_mod} 
\end{figure*}

\subsection{Dynamical classes, masses and disk stability}
In order to characterize the galaxies in our sample, we used a kinematical classification based on the comparison of the morphology of the stars
(showing by the CFHT i-band image) with the morphology of the \halpha\ flux distribution. We classified the galaxies through analysis of the observed
velocity map, the observed velocity dispersion map ($\sigma$, corrected from the instrumental effect), and the velocity residuals map.

On the 10 galaxies of our sample, only two objects present a dynamical structure which is not compatible with rotation and which has not been
successfully fitted by the simple rotating disk model. One object, VVDS-1235, has a complex kinematics with multiple velocity shear and several peak
regions in the $\sigma$-map, and we defined as a \emph{merger} and the other, VVDS-6027, shows negligible velocity shear (defined as \emph{featureless},
with the possibility of being a face-on disk). The remaining eight objects appear compatible with rotating disks and we used the correlation between
the integrated line width, $\sigma_{1D}$, and the maximum velocity dispersion estimated by the best-fitting rotating model to distinguish
rotation-dominated (RD) galaxies from dispersion-dominated (DD) galaxies \citep{Weiner:2006ApJ...653.1027W}. Fig.~\ref{Fig:weiner16} shows that four
galaxies of our sample (VVDS-2331, VVDS-5726, VVDS-4167, and VVDS-1328) are RD disk (open circle) and four others (VVDS-4252, VVDS-4103, VVDS-7106 and
VVDS-6913) are DD disks (filled symbols). Among the DD galaxies we highlighted using a filled square symbol, the galaxy VVDS-6913 which is consistent
with being the relic of a major merging event. And using a filled triangle, the galaxy VVDS-7106 for which the error on the $V_{rot}$ is extremely high.

We also infer the ratios of the maximum velocity over the local velocity dispersion $V_{rot}/\sigma_{0}$ in order to characterize the dominance of the
rotation versus the disordered motions of the gas. We determine the ratio $r_{c}/r_{gas}$ between the radius of the plateau, as estimated by the
rotation modelling, and the radius of nebular emission detected in the SINFONI data to emphasize how close or far to the centre of the object the
maximum of the velocity is reached. We diskuss this two ratios for each galaxy in section~\ref{nature}. The values for $V_{rot}/\sigma_{0}$,
$r_{c}/r_{gas}$ and the classification are shown in Table~\ref{tabkine}.

For objects where the velocity shear is well fitted by the simple rotation model, we compute the dynamical masses assuming that the object is a
circularly rotating disk. Thus the total mass within $r_{c}$ for which the maximum rotational velocity $V_{rot}$ is reached, is approximately described
by:
\begin{equation}
{\rm M_{dyn}}={\rm V_{rot}^{2}}r_{c}/{\rm G}
\end{equation}
where $V_{rot}$ has been correction for inclination effects. Both this asymptotic velocity, and the radius of the turnover ($r_{c}$) are inferred from
our model fits to the observed velocity maps to correct for the the effect of beam-smearing (due to the seeing). The results are given in
Table~\ref{tabkine}.

We also computed the Toomre parameter `$Q$' (Toomre 1964) which is used to quantify the stability of the disk against gravitational collapse. Typically,
values of $Q > 1$ represent dynamically stable systems for which the total baryonic mass can be supported by the observed rotational motion alone, while
values of $Q < 1$ suggest that the observed baryonic total mass exceeds that which can be supported by the observed rotational velocity of the disk,
indicating either that the disk itself is unstable or that non-rotational motions contribute a significant degree of support to the system. Finally, $Q
\sim 1$ represents the case of marginal stability. In order to estimate `$Q$', we considered the ratio of masses within the radius probed by the SINFONI
detections ($r_{disk} = r_{gas}$) and therefore computed the dynamical masses within the gas radius. We do not take the values of the dynamical masses
computed from the rotationally-supported thin disk model ($M_{\rm dyn}$). The total baryonic mass is $M_{\rm disk} = M_{\rm gas} + M_{\star}$. We thus
used the following relation:
\begin{equation}
Q = \frac{V_{rot}^{2}}{G M_{\rm disk}/r_{\rm disk}} = \frac{V_{rot}^{2}}{G (M_{\rm star} + M_{\rm gas})/r_{\rm gas}}
\end{equation}
Results are given in Table~\ref{tabkine}.
\section{Dynamical properties and nature of individual galaxies}
\label{nature} In this Section, we summarize the properties of each galaxy, derived from its spectral energy distribution, mass, star formation rate
and dynamical structure. We use this to provide an insight into the nature and
evolutionary status of each the 10 galaxies in our sample.
\subsection{VVDS-1235}
This galaxy has the lowest redshift of our sample and the lowest stellar mass ($\sim$ 0.6 x $10^{10}\, M_{\odot}$) . It exhibits complex kinematics, and
it is impossible to find a clear centre for this clumpy galaxy. Its $\sigma$-map shows several peaks which seem to match with the knots in the H$\alpha$
intensity map. The peak with the maximum H$\alpha$ flux shows a very high dispersion of $\sim240$ km s$^{-1}$. Our simple rotating disk model did not
succeed in reproducing its complex velocity field; we interpret this galaxy as being a merging system.
\subsection{VVDS-2331}
This galaxy shows a velocity gradient of $\pm50$ km s$^{-1}$ in the south-west/north-east direction (position angle: $-69\lyxmathsym{\textdegree}$).
However, we only cover the central part since the plateau is not reached in our SINFONI observation. Its $\sigma$-map show a peak at $120$ km s$^{-1}$
slightly shifted from the \halpha\ flux peak. However this peak also corresponds to a region with high noise. The rotational disk model fits the
observed inner gradient relatively well, and we found a plateau velocity, corrected from beam smearing, of $V_{rot} \sim 197$ km s$^{-1}$ within 4.8
kpc. We found a virial mass of $\sim$ 6 x $10^{10}\, M_{\odot}$ and a dynamical mass of $\sim$ 0.4 x $10^{10}\, M_{\odot}$, estimated from the $V_{rot}$
inferred by the rotation modelling. It is one of the objects for which we found the lowest value of local velocity dispersion ($<$ 40 km s$^{-1}$),
leading to a $V_{rot}/\sigma_{0}\sim5$. VVDS-2331 also seems to have a large amount of gas ($\mu\sim$ 70\%), a high SFR and a young stellar population.
It is therefore a good candidate for being a cold rotating disk {\bf ($Q \sim 3$)}, where rapid star formation is occurring due to the injection of
large quantities of cold gas.
\subsection{VVDS-6913}
The H$\alpha$-map shows two components, one of which is coincident with the peak of the \halpha\ flux. However, the kinematical centre is
located exactly between the two components, at $\sim$ 8 kpc from the centre of the main blob. It has a smooth velocity gradient and a large peak in the
centre of its $\sigma$-map. The whole velocity map, including the two blobs, is well fitted by the rotation model (plateau velocity of $\sim$
139 km s$^{-1}$ reached at 10.7 kpc and $V_{rot}/\sigma_{0}\sim3$). However the small arms linked to the large peak in the $\sigma$-map reveal
a perturbed rotation. The peak of the dispersion and the arms are still visible in the residual map of the dispersion (see
Fig.~\ref{map3_mod}), which emphasize additional random motions (up to 70 km s$^{-1}$ with a mean of 20 km s$^{-1}$). This morphology can also
be explained by the presence of a relic of a previous merger event, or by being a clumpy disk galaxy. VVDS-6913 has the second highest virial mass ($\sim$ 20 x
$10^{10}\, M_{\odot}$) and dynamical mass ($\sim$ 5 x $10^{10}\, M_{\odot}$) of the galaxies in our sample. It has a reasonable SFR and also a
sizeable stellar mass ($\sim$ 4 x $10^{10}\, M_{\odot}$) and the remaining quantity of gas ($\mu<$ 35\%) indicates that it has already turned a
large amount of its gas into stars. We classify it as a DD rotating disk, since its disk stability appears marginal ($Q \sim 2$).
\subsection{VVDS-5726}
This nearly edge-on galaxy has its H$\alpha$-map and $\sigma$-map peaked at the centre. The general shape of the velocity gradient is well reproduced by
the rotational disk model with a maximum velocity of $V_{rot} \sim 292$ km s$^{-1}$ reached at 1.5 kpc and a local velocity dispersion of $\sim$ 60 km
s$^{-1}$ ($V_{rot}/\sigma_{0}\sim5$). We found a virial mass of $\sim$ 9 x $10^{10}\, M_{\odot}$ and a dynamical mass of $\sim$ 3 x $10^{10}\,
M_{\odot}$ for this object. This galaxy has the highest stellar mass ($\sim$ 4 x $10^{10}\, M_{\odot}$), the lowest gas fraction ($<$ 30\%) and one of
the oldest stellar population (1.26 Gyr) of all the objects in our sample. Also taking into account its moderate SFR, we believe that this galaxy has a
rotating disk of cold gas ($Q \sim 3$) and it is at a later evolutionary state than most of the objects in our sample.
\subsection{VVDS-4252}
The resolved velocity structure present a smoothly varying shear along the major axis, including the beginning of a plateau at a radial
velocity of $\sim\pm54$ km s$^{-1}$. There is a very good agreement between the best-fit rotating disk model,  with a maximum $V_{rot} \sim 130$ km s$^{-1}$ reached at 4.5 kpc, and the observed rotation curve (see
Fig.~\ref{map4_mod}). The residual maps show very low mean and rms values
(see Table~\ref{tabmodel}). However, the velocity dispersion map is broadly peaked at the centre of the galaxy and the high value local
velocity dispersion ($\sim$ 100 km s$^{-1}$) probably indicates that the galaxy does not have a dynamically cold rotating disk of ionized gas, due
to significant disordered motion in the gas ($V_{rot}/\sigma_{0}\sim1$). We also found that the virial mass ($\sim$ 14 x $10^{10}\, M_{\odot}$) is much higher than the
dynamical mass ($\sim$ 2 x $10^{10}\, M_{\odot}$). It has a young stellar population, a very high gas fraction ($\sim$ 87\%) and seems to be
undergoing a strong episode of star formation. VVDS-4252 is probably consistent with a rotation in a heated disk (DD unstable rotating disk ($Q \sim 0.4$)).
\subsection{VVDS-4103}
This galaxy has a velocity gradient which can be fitted by a simple rotating disk model, and presents a clear peak in its $\sigma$-map which corresponds to the maximum of the H$\alpha$ flux distribution. The best-fit disk model indicates a maximum velocity of
$V_{rot} \sim 118$ km s$^{-1}$ km/s, reached at 8.5 kpc and a local velocity dispersion of $\sim$ 58 km s$^{-1}$. Hovewer, the peak in the residual dispersion map (see Fig.~\ref{map4_mod}) and the low ratio $V_{rot}/\sigma_{0}\sim2$ indicates the presence of non-negligeable random motions in the gas. We also found that the virial mass ($\sim$ 12 x $10^{10}\, M_{\odot}$) is much higher than the
dynamical mass ($\sim$ 3 x $10^{10}\, M_{\odot}$). It has a extremely high SFR and gas fraction. VVDS-4103 must be experiencing a very strong burst of star formation, which might be the cause of non-negligible random motions of the gas peaked at the centre of the object. These galaxy properties are quite similar to the properties of VVDS-4252 and we also classify it as a DD unstable rotating disk ($Q \sim 0.3$).
\subsection{VVDS-4167}
This galaxy exhibits a smoothly varying shear along its major axis and has a clear peak in its $\sigma$-map corresponding to the maximum of
the H$\alpha$ flux distribution which is slightly shifted from the centre of the object. The velocity field is relatively well fitted by our
simple rotational disk model with a maximum velocity of $V_{rot} \sim 257$ km s$^{-1}$ km/s, reached at 10.8 kpc, and a local velocity
dispersion of $\sim$ 54 km s$^{-1}$. This leads to a $V_{rot}/\sigma_{0}\sim5$ showing the dominance of the rotation over the random motions of
the gas, even if the peak near the centre is also visible in the residual of the dispersion map (see Fig.~\ref{map4_mod}). We found a virial
mass of $\sim$ 26 x $10^{10}\, M_{\odot}$ and a dynamical mass of $\sim$ 17 x $10^{10}\, M_{\odot}$ for this object. It also has a sizeable
and very young stellar population, a reasonable SFR and has turned slightly more than half of its gas into stars. We believe therefore that this
galaxy can be consistent with a cold rotating disk ($Q \sim 8$).
\subsection{VVDS-7106}
This galaxy seems to have a low velocity shear (see Fig.~\ref{map1_fig}). As with VVDS-2331, we cover only the central part of this object and therefore the plateau has not been reached. The rotational disk model fits  the observed
inner gradient relatively well and gives an estimated maximum velocity of $V_{rot} \sim 105$ km s$^{-1}$ km/s, reached at 9.4 kpc, and a local velocity
dispersion of $\sim$ 82 km s$^{-1}$. We also found that the virial mass ($\sim$ 4 x $10^{10}\, M_{\odot}$ is similar to the
dynamical mass ($\sim$ 2 x $10^{10}\, M_{\odot}$), and that it has a ratio of $\frac{r_{c}}{r_{gas}} >$ 1. We measure a low ratio of $V_{rot}/\sigma_{0}\sim1$, showing the dominance of random motions in the gas as also indicted by the high local velocity dispersion. VVDS-7106 has the lowest stellar mass in our sample  and an extremely young stellar population. More than 60\% of its gas has been turned into star and it has an average SFR. Taking into account all of its properties, we classify it as a DD rotating disk despite its marginal stable disk ($Q \sim 1$).
\subsection{VVDS-6027}
This object has a young stellar population with an average stellar mass of $\sim$ 1 x $10^{10}\, M_{\odot}$, a low gas fraction of ($<$ 35\%) and a small SFR. It presents an \halpha\ flux distribution with a peak located at its centre and a low surface brightness emission tail seen also in the  stellar population image (see Fig.~\ref{map2_fig}).
This galaxy is relatively well-resolved but shows no strong evidence for spatially resolved velocity structure. We classify it as featureless (perhaps consistent with being a face-on galaxy).  
\subsection{VVDS-1328}
As for VVDS-5726, this nearly edge-on galaxy has a velocity gradient which is well reproduced by the rotational disk model. However,
the peak seen in the $\sigma$-map is slightly shifted from the peak in the \halpha\ flux map (see Fig.~\ref{map1_fig}). The best-fit disk model
indicates a maximum velocity of $V_{rot} \sim 182$ km s$^{-1}$ km/s, reached at 0.5 kpc, and a local velocity dispersion of $\sim$ 36 km
s$^{-1}$ ($V_{rot}/\sigma_{0}\sim5$). We found a virial mass of $\sim$ 3 x $10^{10}\, M_{\odot}$ and a dynamical mass of $\sim$ 0.4 x $10^{10}\, M_{\odot}$ for this
object. This galaxy has an average stellar mass of $\sim$ 1 x $10^{10}\, M_{\odot}$, a low gas fraction of ($<$ 35\%) and a small SFR. We
believe that this `mature' galaxy has a rotating disk of cold gas ($Q \sim 6$).
%
\begin{table*}
\caption{Kinematical properties}
\begin{tabular}{ccccccccc}
\hline 
Galaxy  & $v_{shear}$ & $V_{rot}/\sigma_{0}$ & $M_{vir}$ (1) & $M_{dyn}$ (1) & $\frac{r_{c}}{r_{gas}}$  & $Q$ (2) & dyn. class (3)\tabularnewline
\hline 
VVDS-1235  & 92$\pm$13 & -  & $4\pm2$  & -  & - & - & merger.\tabularnewline
VVDS-2331  & 99$\pm$17 & $5.1\pm2.9$  & $6.4\pm2.9$  & $0.4\pm0.7$  & $0.52\pm0.87$  & $3.3\pm0.9$ & RD  \tabularnewline
VVDS-6913  & 139$\pm$7 & $3.2\pm0.8$  & $20\pm6.5$  & $5\pm3$  & $0.53\pm0.15$ & $2.4\pm0.6$ &  DD \tabularnewline
VVDS-5726  & 95$\pm$9 & $4.9\pm0.5$  & $8.7\pm0.3$  & $3\pm4$  & $0.22\pm0.3$ & $3.3\pm0.5$ &  RD \tabularnewline
VVDS-4252  & 54$\pm$9 & $1.3\pm0.1$  & $14\pm8.7$  & $2\pm1$ & $0.42\pm0.19$ & $0.4\pm0.7$ &  DD \tabularnewline
VVDS-4103  & 83$\pm$16 & $2.0\pm0.7$  & $12\pm13$  & $3\pm2$  & $0.63\pm0.23$ & $0.3\pm0.8$ & DD \tabularnewline
VVDS-4167  & 140$\pm$8 & $4.8\pm0.8$  & $26\pm2.2$  & $17\pm5$  & $0.65\pm0.13$ & $7.5\pm0.5$ & RD   \tabularnewline
VVDS-7106  & 12$\pm$14  & $1.3\pm0.4$ & $4.1\pm0.7$ & $2\pm3$ & $1.22\pm0.58$ &  $1.4\pm1.0$ & DD  \tabularnewline
VVDS-6027  & $<$10$\pm$11  & - & $1\pm0.5$ & - & - &  - &  f.l.\tabularnewline
VVDS-1328  & 58$\pm$8 & $5.1\pm0.7$  & $3.4\pm1.0$  & $0.4\pm3$  & $0.07\pm0.56$ & $5.7\pm0.7$  & RD \tabularnewline
\hline
\end{tabular}

\begin{raggedright}
The columns are as follows: (1) $10^{10}\, M_{\odot}$; (2) Toomre parameter (Toomre 1964); (3) dynamical classification: RD for \emph{rotation-dominated} rotating disk, DD for \emph{dispersion-dominated} rotating disk, and f.l. for \emph{featureless}.
\par\end{raggedright}

\label{tabkine}
\end{table*}
\section{diskussion} 
\label{relations}
Here, we study and diskuss the relations between the kinematical and physical properties of the ionized gas and 
the physical properties of the stellar population for the 10 galaxies of our sample.
\subsection{Relations between the recent-to-past star formation rate and other properties}
We calculated the recent-to-past star formation rate ratio ($SFR_{corr}/SFR_{sed}$), which is the ratio between the star formation rate
estimated from the ionized gas (using the \halpha\ emission) and the one estimated from the best-model SED fitting (both of which have been dereddened).
$SFR_{sed}$ is averaged on a timescale ten times longer than the $SFR_{corr}$, which is an indication of the instantaneous star formation.
Given these different timescales, the ratio between the two estimates gives information on the presence of recent burst of star formation
above the standard continuous declining star formation. We investigated, thus, the correlation between the recent-to-past star formation rate
ratio ($SFR_{corr}/SFR_{sed}$) and the gas mass fraction and the stellar mass.
\begin{figure}
\begin{centering}
\includegraphics[width=0.9\columnwidth]{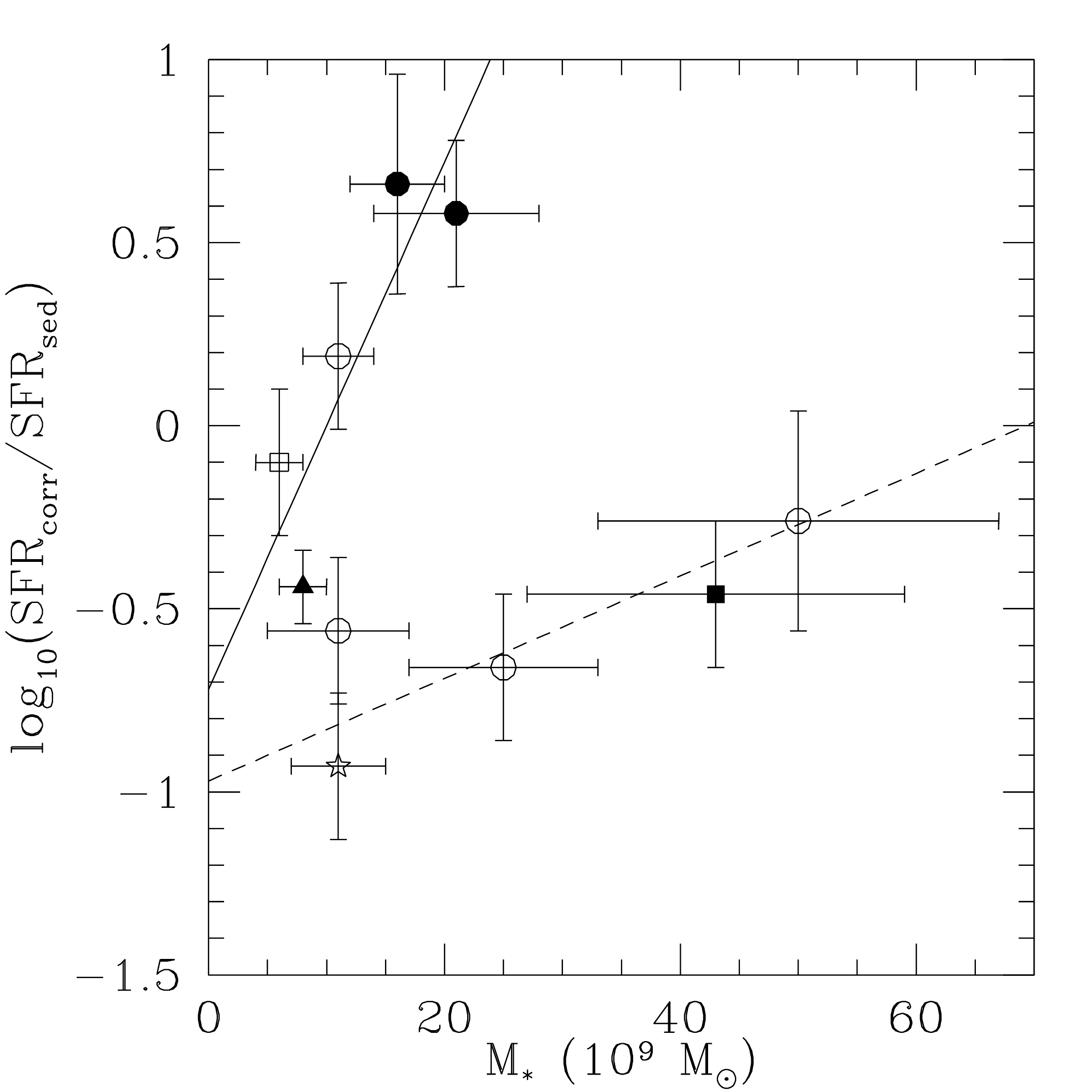}\\ 
\par\end{centering}
\caption{Relation between the recent-to-past star formation rate ratio and the gas mass fraction. The symbols give
the dynamical classification as follows:  \emph{open circle}: RD rotating disks; \emph{filled symbols}: DD rotating disks, with in particular \emph{filled triangle}: VVDS-7106, and \emph{filled square}: VVDS-6913; \emph{star}: featureless. We found two classes (see text) and a least-square linear fit is performed for each of them.}
\label{sfr_mu_age}
\end{figure}
\subsubsection{Relations between the recent-to-past star formation rate ratio and the gas mass fraction}
In Fig.~\ref{sfr_mu_age}, we plot the recent-to-past SFR ratio vs.  the  gas mass fraction for the 10 galaxies of our sample.
We found that only galaxies with a significant amount of gas ($\mu\sim >$0.6), are undergoing episodes of strong star formation (bursts).
From Fig.~\ref{sfr_mu_age}, a correlation between the recent-to-past SFR ratio and the gas mass fraction, given that we divide the sample into two classes is therefore clearly seen. The first one include 
the three objects which have the smallest
recent-to-past SFR ratio. These three objects are: the face-on/featureless VVDS-6027 (star symbol),  the DD rotating galaxy VVDS-7106 (filled triangle), and the RD rotating galaxy VVDS-4167 (open cirle), and therefore are not showing evidence of bursts of star formation. This first class seems to follow the relation (least square fit):
\[log_{10}(SFR_{corr}/SFR_{sed})=1.38(\pm0.45)\times\mu-1.40(\pm0.24),\]
and the second class, all the remaining objects, seems to follow the relation:
\[
log_{10}(SFR_{corr}/SFR_{sed})=1.77(\pm0.01)\times\mu-1.03(\pm0.01).\]
Can these two classes be related to underlying kinematical properties ? We find indeed that the
ratio of the kinematical radius to the gas radius ($r_c/r_{gas}$, see Table~\ref{tabkine}) has a mean value
of $0.40\pm0.01$ for the galaxies with a high recent-to-past SFR ratio, and $0.94\pm0.01$ for the three others of the first class. It appears, thus, that the rather intuitive relation between the amount
of recent star formation and the amount of ionized gas depends strongly
on kinematics. `Slow rotator' (intermediate redshift galaxies with a $r_c/r_{gas} > 0.6$) appear less efficient in converting a high mass fraction of ionized gas into a burst of star formation.
This will have to be confirmed and refined with better data and more objects.
\begin{figure}
\begin{centering}
\includegraphics[width=0.9\columnwidth]{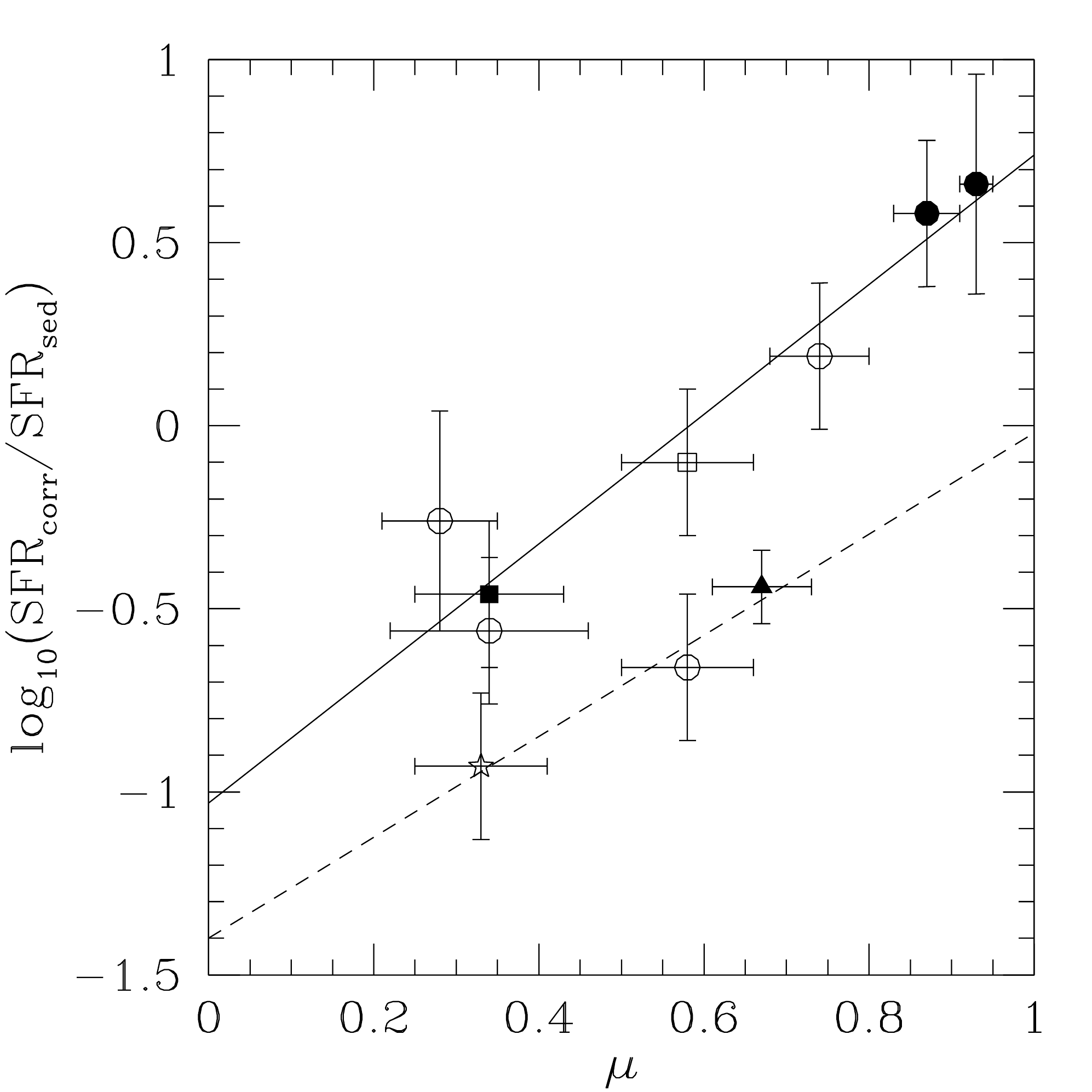} 
\par\end{centering}
\caption{Relation between the recent-to-past star formation rate ratio and the logarithm of the stellar mass. The symbols are the same than in Fig.~\ref{sfr_mu_age}}
\label{sfr_mass}
\end{figure}
\subsubsection{Relations between the recent-to-past star formation rate ratio and the stellar mass}
In Fig.~\ref{sfr_mass}, we plot the recent-to-past SFR ratio vs. the stellar mass for the 10 galaxies of our
sample. We found that the galaxies of our sample need higher stellar masses to reach higher recent star formation
rates. The recent star formation to stellar mass efficiency is moreover significantly stronger for half of our
sample. From Fig.~\ref{sfr_mass}, we see two possible correlations between the recent-to-past SFR ratio and the
stellar mass. The first class includes objects dominated by recent star formation ($SFR_{corr}/SFR_{sed}>1$) : the
merger object and all DD galaxies, except VVDS-6913 with the following relation (least squares fit):
\[log_{10}(SFR_{corr}/SFR_{sed})=0.072(\pm0.05)\times M_{\star}(10^{9}M_{\odot})\]
\[\hspace{3.8cm}-0.72(\pm0.4)\]
The second class
includes objects dominated by past star formation ($SFR_{corr}/SFR_{sed}<1$): the featureless object and all
RD galaxies, except VVDS-2331, seems to follow the relation (least squares fit) 
\[log_{10}(SFR_{corr}/SFR_{sed})=0.014(\pm0.03)\times M_{\star}(10^{9}M_{\odot})\]
\[\hspace{3.8cm}-0.97(\pm0.7)\] 
The star formation
to stellar mass efficiency is moreover stronger for objects experiencing a burst of recent star formation, and
appears also to be related to the dominance of random motions in the gas. We indeed notice that the ratio
of the maximum velocity to the velocity dispersion ($V_{rot}/\sigma_0$; see Table~\ref{tabkine}) is 
different for the rotating disks galaxies inside the two classes: it has a mean value of $2.42\pm0.5$ for the
first class, and $4.5\pm0.3$ for the second one. 
We thus propose to refine our classification: disk galaxies of the first family would be called "hot/perturbed
rotators", while the disk galaxies of the second family would be `cold rotators'. Only hot rotators and
objects dominate by random motions, like the merger seem to be able to experience a recent burst. The separation between cold rotators
and hot/perturbed rotators would then be $V_{rot}/\sigma_0\approx4$. This also will have to be confirmed and
refined with better data and more objects.
\subsection{The stellar mass Tully-Fisher relation} 
\begin{figure}
\begin{centering}
\includegraphics[width=1\columnwidth]{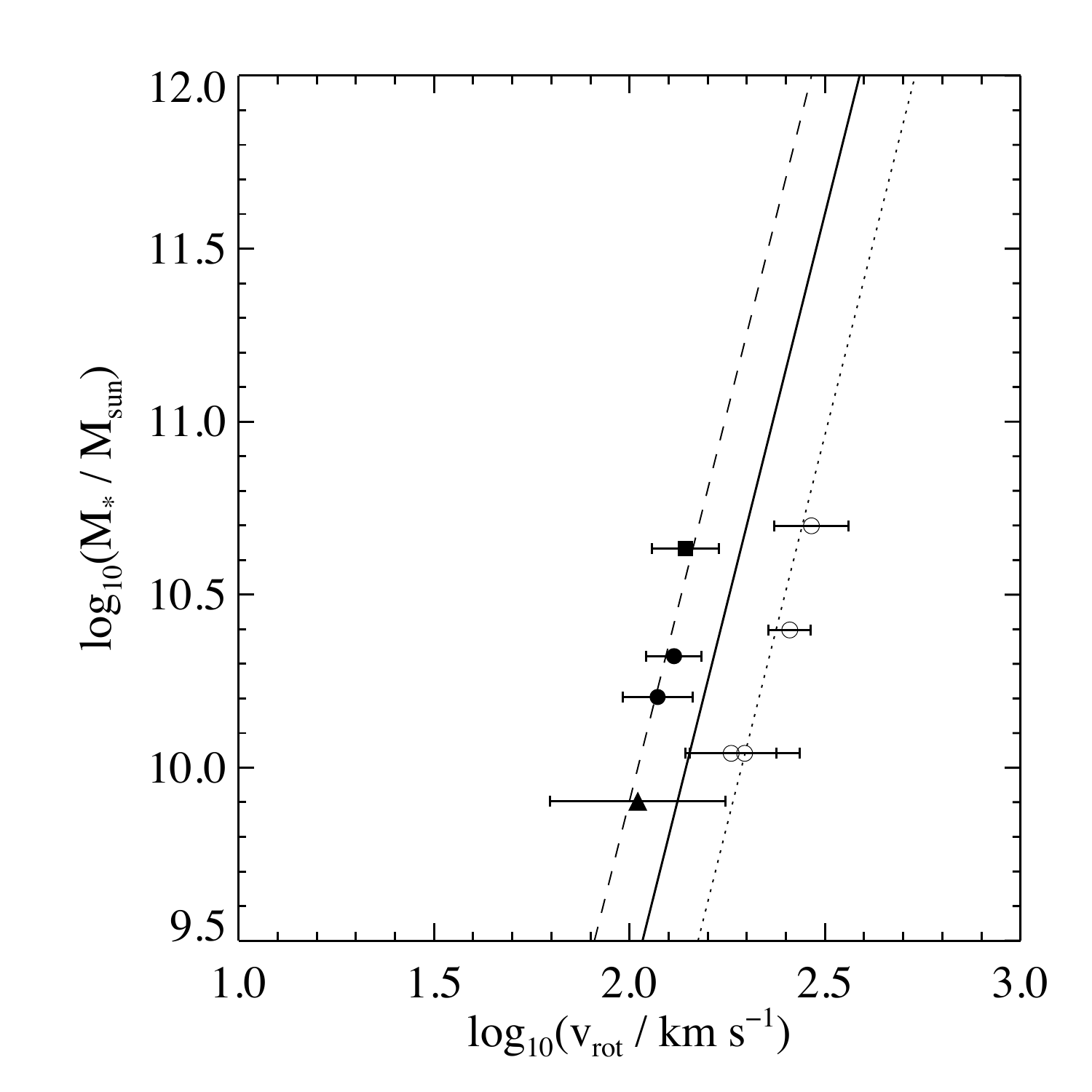}\\
\includegraphics[width=1\columnwidth]{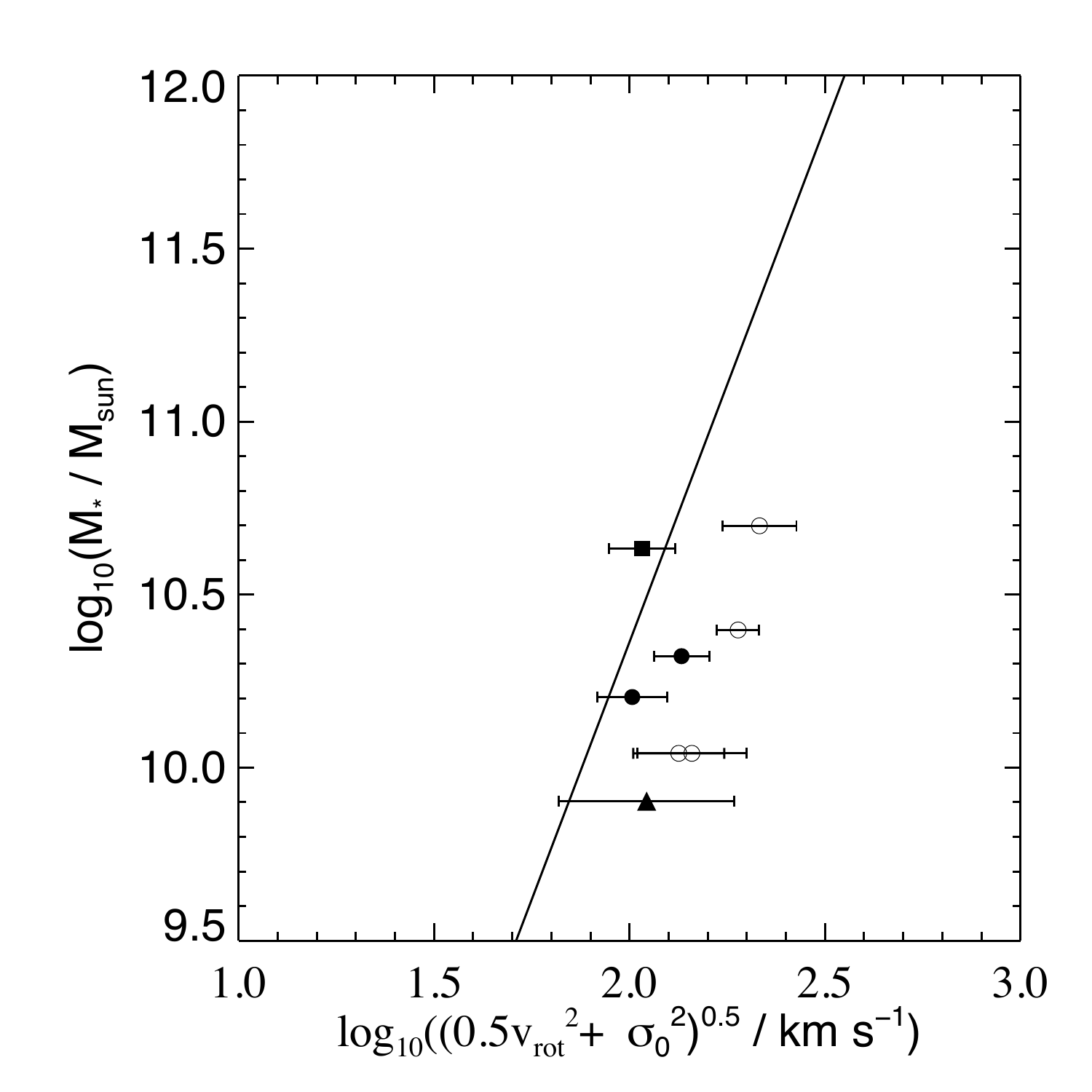}
\par\end{centering}
\caption{The stellar mass Tully-Fisher relation at 1.2$<$z$<$1.5, using the maximum velocity (top panel) and using the $S_{0.5}$ index (bottom panel; $(0.5 V_{rot}^2 + \sigma_{0}^2)^{0.5}$). In the top panel, the solid line is the local relation from \citep{2001ApJ...550..212B}, while in the bottom panel it is the z$\sim$1 relation from \citep{Kassin:2007ApJ...660L..35K}. The symbols are \emph{open cirle}: RD rotating disks and \emph{filled symbols}: DD rotating disks with in particular \emph{filled triangle}: VVDS-7106, and \emph{filled square}: VVDS-6913.}
\label{Fig:tully}
\end{figure}
In Fig.~\ref{Fig:tully}, we show the Tully-Fisher relations for the 1.2$\ltapprox$z$\ltapprox$1.5 galaxies showing evidence of rotation. In the top
panel, we plotted the stellar mass versus the maximum velocity. As can be seen, there is a clear separation between the RD rotating disks (open circle)
from the DD rotating disks (filled symbols). However, both groups of objects suggest a good correlation between the stellar mass and the rotation
maximum speed similar to that seen in the local universe (sold line). By assuming the local slope of the Tully-Fisher relation from
\citep{2001ApJ...550..212B}, we found a best fitting zero point of $-0.29 \pm 0.25$ for the RD galaxies and $0.90 \pm 0.43$ for the DD galaxies.

In the bottom panel of Fig.~\ref{Fig:tully}, we replace the maximum velocity by the $S_{0.5}$ index ($0.5 V_{rot}^2 +\sigma_{0}^2)^{0.5}$), which allows
the disordered motions in the gas to be taken into account \citep{Kassin:2007ApJ...660L..35K}. The separation between the RD rotating disks and the DD
rotating disks is less obvious while adding the disordered motions. It seems that we also detect an evolution of the relation since z$\sim$1 (which is
shown as the solid line in Fig.~\ref{Fig:tully}).
%
\subsection{Comparison with theoretical studies, with results from other intermediate- and high-redshift samples} 
\begin{figure}
\begin{centering}
\includegraphics[width=1\columnwidth]{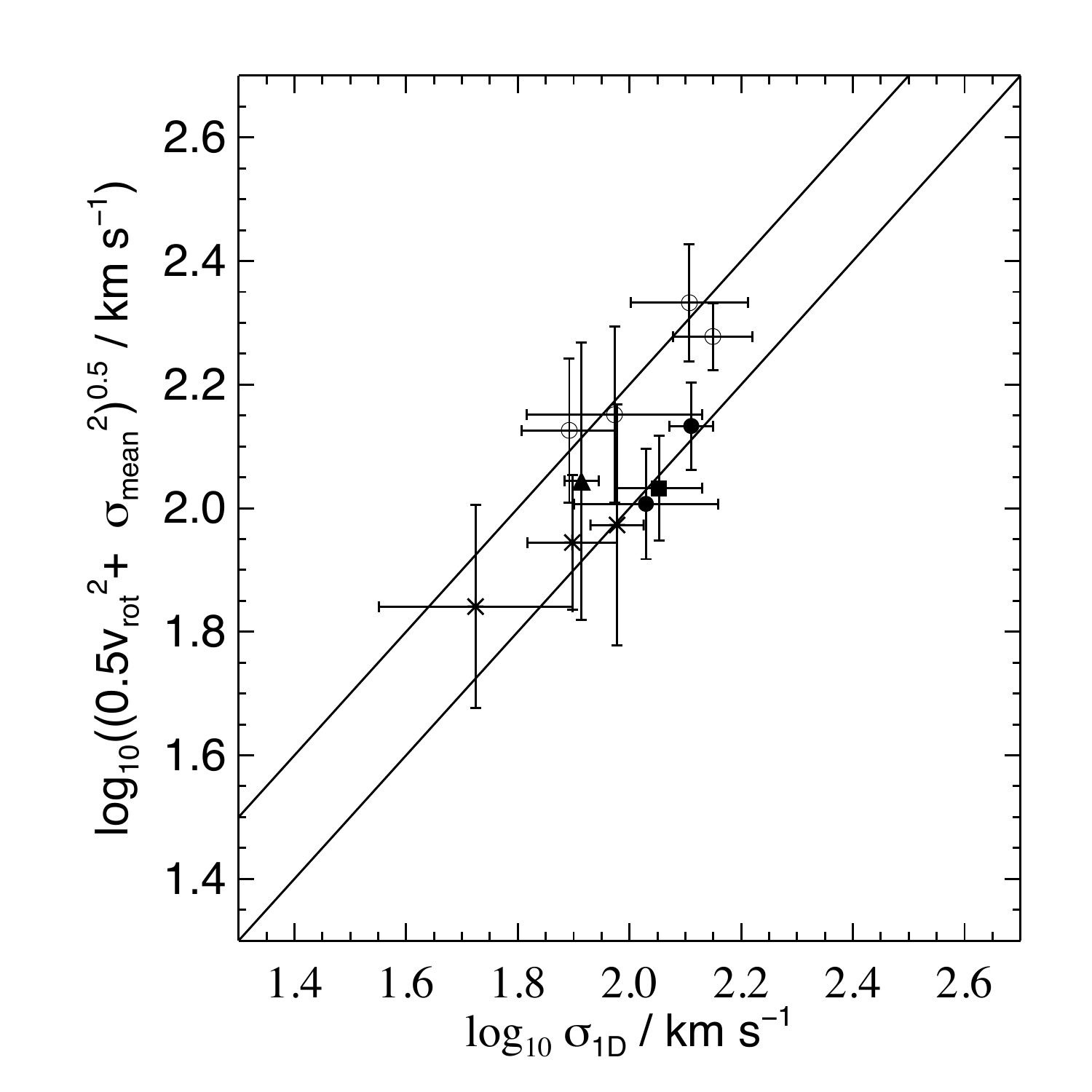}
\par\end{centering}
\caption{Combined velocity (\whalf\ ) versus integrated linewidth velocity dispersion. The diagonal lines are the same as in \citet{Wright:2007ApJ...658...78W}, i.e.\ a $1:1$ line and the \citet{1997MNRAS.285..779R} $\sigma = 0.6 V_c$ line ($V_c$ is the circular velocity). Linewidth $\sigma_{1D}$ and \whalf\ are correlated; the 0.5 pre-factor makes the combined velocity width a better estimate of velocity dispersion, so that the correlation is tighter and the galaxies closer to the $1:1$ line. The symbols are  \emph{open cirle}: RD rotating disks and \emph{filled symbols}: DD rotating disks with in particular \emph{filled triangle}: VVDS-7106, and \emph{filled square}: VVDS-6913;  \emph{asterisk}: $z \sim 3$ galaxies from \citet{Lemoine-Busserolle:2009}}
\label{Weiner2006_fig19}
\end{figure}

Previous studies at $z > 2$ have shown the presence of large quantity of gas in comparison to the stellar population in high-redshift galaxies \citep{2009arXiv0901.2930L,Lemoine-Busserolle:2009}. We found that six out of then galaxies of our sample have a gas fraction higher than 0.50. The typical $v/\sigma$  for our sample is similar to the one found for the SINS galaxies ($\sim 2 - 4$; \citet{Forster:2006ApJ...645.1062F,Genzel:2006Natur.442..786G,Genzel:2008ApJ...687...59G}), but more higher than the values found by \citet{Lemoine-Busserolle:2009} ($\sim 0.4-1.5$) and \citet{2009arXiv0901.2930L} galaxies at $z\sim 2$ ($\sim 0.8$). \citet{Lemoine-Busserolle:2009} found in their sample, 3 galaxies showing the presence of rotation. Fig~\ref{Weiner2006_fig19} plots the combined velocity
scale \whalf\ against the 1D linewidth \sigoned\ for our sample together with this three galaxies at $z \sim 3-4$.  It  should give an idea of the evolution of the dynamical mass which take also into account the contribution from the random motions in the gas. Fig~\ref{Weiner2006_fig19} shows an increase of the dynamical mass for $z \sim 3-4$ to $z \sim 1.3-1.5$ for the DD disks. 

Despite the low resolution of the morphological data (CFTHLS I-band images) which shows that, except VVDS-1235, all the galaxies seems to be disk like objects with a central bulge, the IFU data obtained on the sample of the 10 star-forming galaxies at 1.0$\ltapprox$z$\ltapprox$1.5 have confirmed the presence of a velocity gradient resembling that expected for an internally rotating system, but has also reveals the presence of non-negligible random motions in the gas. Recent studies at the same redshift have found similar results.  \citet{Bournaud:2008A&A...486..741B} have study a clumpy galaxy at $z=1.6$, which was previously classified as an on-going merger. Their study reveals that the kinematics properties (the large-scale rotation with $V_{rot} \sim 100$ km s$^{-1}$), the stellar properties (young age, stellar mass range) and the properties of the gas are very similar to those of the galaxies of our sample (in particular VVDS-6913). Their numerical model shows that a such turbulent rotation disk,  results more from the evolution of an unstable gas-rich disk galaxy than from a merging event. \citep{Wright:2007ApJ...658...78W,Wright:2009ApJ...699..421W,Bournaud:2008A&A...486..741B}

Five of the rotating disk of our sample have a $Q > 2$, which indicates the stability of their disk. Can this large stable disk systems evolve in present-day thin disks ? We already note that the disk galaxies in our sample (except the two galaxies VVDS-1328 and VVDS-5726 which already have similar properties to that of the present-day local spirals) possess a large disk of gas and younger stars (age < 1 Gyr), which can further grow in mass by continued accretion. The lack of rotation observed in this objects in comparison to the thin local disk is nowadays well known and taken into recent plausible formation and evolution of high-redshift thick disk models. These scenarios can predict that on the largest scales a velocity gradient tracing a rotation can be observed with high velocity dispersion.
They also shown that before  $z\sim2$ the hot mode of accretion dominates, but after  $z\sim 2$, the galaxy has a thin and extended  disk component with $Q \simeq 1.5-2$, which indicates that the disk is marginally stable. They indicate also that after these earlier stages the galaxy enters a slow accretion phase and the disk evolves quiescently until $z=0$ \citep{keres05,DekelBirnboim06,Ocvirk08,Dekel09a}. 
The results presented in this paper support the hypothesis that stable gas-rich disks seen at intermediate and high-redshifts may internally evolve into present-day spirals. 
\section{Conclusions} 
\label{summary} 
We have presented the 2D kinematics and the physical properties of a sample of ten  star-forming disk galaxies at 1.0$\ltapprox$z$\ltapprox$1.5. 
Among these objects, three (VVDS-4252, VVDS-4103 and VVDS-4167) are undergoing a strong burst of star formation. We found mainly four   kinematical types in our sample which are: one merger (VVDS-1235); one featureless (or face-on) galaxy (VVDS-6027); four DD rotating disks $-$ VVDS-4252, VVDS-7106, VVDS-4103 and VVDS-6913 which is also consistent of being the relic on a major merging event (to which the disk system seems to have survived) or a good candidate for being a clumpy galaxy; and finally four other galaxies are RD disks $-$ VVDS-2331, VVDS-4167, including VVDS-1328 and VVDS-5726 which are pure rotationally supported disks.
These two rotating disks achieve a maximum velocity  of $\sim$ 180-290 km s$^{-1}$ km/s within $\sim$ 0.5-1 kpc, similar to local spirals with thin disk \citep{1999ApJ...523..136S}. Regarding most of the DD rotating disk, they display a plateau velocity range of 105-257 km s$^{-1}$ km/s, certainly underestimated due to beam smearing. However, their plateau radii (4.5-10.8 kpc) derived from our rotating disk model are significantly higher than those derived for pure rotating disks and local spiral galaxies. 
We did not find any trace of AGN in the 10 objects. 
The galaxies of our sample have a relatively young stellar population ($<$ 1 Gyr) and possess a range of stellar mass of 0.6-5 $10^{10}\, M_{\odot}$. In addition, most of them have not yet converted the majority of their gas into stars (six galaxies have their gas fraction $>$ 50 per cent). Therefore, the galaxies which already have a stable disk (six of them) would have their final stellar mass similar to the present-day spirals, to which these rotating systems can be seen as precursors.
We also investigated the  stellar mass Tully-Fisher relation at 1.2$\ltapprox$z$\ltapprox$1.5 and found a scatter between the RD disk galaxies and the DD disk galaxies. We also found for this two groups of objects a change of the zero point in comparison with the stellar mass Tully-Fisher relation in the local universe,  but this is speculative considering the statistics of our sample.

Although, we have presented results on the dynamical structure and the physical properties of a small sample of star-forming galaxies at 1.0$\ltapprox$z$\ltapprox$1.5, we have been able to investigate the dynamical type and the physical properties of late-type objects at this intermediate redshifts. Increasing the samples at different redshift ranges for which near-infrared IFS data and multi-wavelength broad-band photometric data can be obtained, will definitely develop our understanding of the dynamical characteristics and nature of various systems (mergers, rotating disks, etc.) and therefore will lead us to understand how galaxies have evolved to match the present-day Hubble sequence.  

\section*{Acknowledgments}

We would like to thank Aprajita Verma and Andy Bunker for helpful
diskussions. We are very grateful to the VLT Observatory for accepting
this programme. The authors also thank Markus Hartung for help obtaining
the observations during the observing runs and Sebastian S\'anchez for providing the code to create the kinematics maps from the 3D cube. The anonymous referee is greatly acknowledged for providing useful and constructive comments. The author wish to recognize
and acknowledge the significant contribution the VVDS collaboration
has done by providing the targets. In particular Thierry Contini who
helped greatly with the targets selection. Part of this work was supported
by the Marie Curie Research Training Network \textit{Euro3D; contract
No. HPRN-CT-2002-00305}. M. Lemoine-Busserolle is supported by the Gemini Observatory, which is operated by the Association of Universities for Research in Astronomy, Inc., on behalf of the international Gemini partnership of Argentina, Australia, Brazil, Canada, Chile, the United Kingdom, and the United States of America.

\end{document}